\documentclass[%
 twocolumn,
 amsmath,amssymb,
 aps,
prc
]{revtex4-2}
\usepackage{graphicx}

\usepackage{dcolumn}
\usepackage{bm}
\usepackage{xcolor}
\usepackage{ulem}
\usepackage{gensymb}

\begin{document}

\title{Measurements of the $^{96}$Zr($\alpha$,n)$^{99}$Mo cross section for astrophysics and applications}
\author{G. Hamad}
\email{gh824514@ohio.edu}
\affiliation{Institute of Nuclear \& Particle Physics, Department of Physics \& Astronomy, Ohio University, Athens, Ohio 45701, USA}
\author{K. Brandenburg}
\affiliation{Institute of Nuclear \& Particle Physics, Department of Physics \& Astronomy, Ohio University, Athens, Ohio 45701, USA}
\author{Z. Meisel}
\email{meisel@ohio.edu}
\affiliation{Institute of Nuclear \& Particle Physics, Department of Physics \& Astronomy, Ohio University, Athens, Ohio 45701, USA}
\author{C.R. Brune}
\affiliation{Institute of Nuclear \& Particle Physics, Department of Physics \& Astronomy, Ohio University, Athens, Ohio 45701, USA}
\author{D.E. Carter}
\affiliation{Institute of Nuclear \& Particle Physics, Department of Physics \& Astronomy, Ohio University, Athens, Ohio 45701, USA}
\author{D.C. Ingram}
\affiliation{Institute of Nuclear \& Particle Physics, Department of Physics \& Astronomy, Ohio University, Athens, Ohio 45701, USA}
\author{Y. Jones-Alberty}
\affiliation{Institute of Nuclear \& Particle Physics, Department of Physics \& Astronomy, Ohio University, Athens, Ohio 45701, USA}
\author{T.N. Massey}
\affiliation{Institute of Nuclear \& Particle Physics, Department of Physics \& Astronomy, Ohio University, Athens, Ohio 45701, USA}
\author{M. Saxena}
\affiliation{Institute of Nuclear \& Particle Physics, Department of Physics \& Astronomy, Ohio University, Athens, Ohio 45701, USA}
\author{D. Soltesz}
\affiliation{Institute of Nuclear \& Particle Physics, Department of Physics \& Astronomy, Ohio University, Athens, Ohio 45701, USA}
\author{S.K. Subedi}
\affiliation{Institute of Nuclear \& Particle Physics, Department of Physics \& Astronomy, Ohio University, Athens, Ohio 45701, USA}
\author{A.V. Voinov}
\affiliation{Institute of Nuclear \& Particle Physics, Department of Physics \& Astronomy, Ohio University, Athens, Ohio 45701, USA}

\begin{abstract}

The reaction $^{96}$Zr($\alpha$,n)$^{99}$Mo plays an important role in $\nu$-driven wind nucleosynthesis in core-collapse supernovae and is a possible avenue for medical isotope production. Cross section measurements were performed using the activation technique at the Edwards Accelerator Laboratory. Results were analyzed along with world data on the $^{96}{\rm Zr}(\alpha,n)$ cross  section and $^{96}{\rm Zr}(\alpha,\alpha)$ differential cross section using large-scale Hauser-Feshbach calculations. We compare our data, previous measurements, and a statistical description of the reaction. We find a larger cross section at low energies compared to prior experimental results, allowing for a larger astrophysical reaction rate. This may impact results of core-collapse supernova $\nu$-driven wind nucleosynthesis calculations, but does not significantly alter prior conclusions about $^{99}{\rm Mo}$ production for medical physics applications. The results from our large-scale Hauser-Feshbach calculations demonstrate that phenomenological optical potentials may yet be adequate to describe $(\alpha,n)$ reactions of interest for $\nu$-driven wind nucleosynthesis, albeit with regionally-adjusted model parameters.

\end{abstract}

\maketitle
\section{\label{sec:level1} Introduction}
The reaction $^{96}$Zr($\alpha$,n)$^{99}$Mo plays an important role in $\nu$-driven wind nucleosynthesis in core-collapse supernovae (CCSN)~\cite{bliss2020} and has been suggested as an accelerator-based production mechanism for $^{99}$Mo~\cite{Hagi18}. The former may help explain the origin of the elements from roughly strontium to silver, while the latter may provide a route to the medical isotope $^{\rm 99m}$Tc that does not require highly-enriched uranium (HEU). Thus, there is considerable interest in a precise determination of the $^{96}{\rm Zr}(\alpha,n)$ cross section for both nuclear astrophysics and nuclear applications.

Astronomical observations of metal poor stars display a decoupling between the elemental abundance pattern for elements in the strontium to silver region, traditionally known as the first rapid neutron-capture ($r$-)process peak, from the abundance pattern for the remainder of the $r$-process~\cite{Mont07,Saka18}. Neutron-rich $\nu$-driven winds of CCSN are a possible contributor to the first $r$-process peak region. The hot protoneutron star produced in core collapse cools via $\nu$ emission, these $\nu$ reheat the supernova shock, and, for neutron-rich conditions, drive nucleosynthesis via $(\alpha,n)$ reactions~\cite{Beth85,Woos92}. The sensitivity of this weak $r$-process nucleosynthesis (also referred to in the literature as the $\alpha$-process and the charged-particle reaction process~\cite{Bliss_2018}) to individual reaction rates can depend on the astrophysical conditions. However, $^{96}{\rm Zr}(\alpha,n)$ generally has a significant influence on nucleosynthesis calculation results as it is usually the main reaction pathway beyond proton-number $Z=40$~\cite{bliss2020}.

Based on the $\approx$2-5~GK temperature range in which $(\alpha,n)$ reactions drive weak $r$-process nucleosynthesis~\cite{Bliss_2017}, $^{96}{\rm Zr}(\alpha,n)$ cross section data are necessary for $\alpha$ laboratory energies in the range $E_{\alpha}\approx5.4-10.9$~MeV. While three measurement results are available near 10~MeV, the stacked-target activation results of Refs.~\cite{Chow95,Vill20} are discrepant with the precision single-target activation results of Ref.~\cite{Kiss21}. Below this beam energy, experimental constraints are only provided by Ref.~\cite{Kiss21}, where data extend down to 6.48~MeV, and therefore the astrophysical reaction rate is essentially exclusively constrained by this single data set. As such, confirmatory data in the energy region of interest are desired.

In nuclear medicine, $^{\rm 99m}{\rm Tc}$ is an important medical isotope for single photon emission computed tomography (SPECT). It is generally produced from HEU targets, where $^{99}{\rm Mo}$ is extracted from the target and $^{\rm 99m}{\rm Tc}$ is subsequently produced on-site with a $^{99}{\rm Mo}/^{\rm 99m}{\rm Tc}$ generator~\cite{Bosc19}. However, the global move away from HEU reactors has threatened this line of production and incentivized the development of accelerator-based production routes~\cite{IAEA2013}. The majority of these approaches require a highly-enriched $^{100}{\rm Mo}$ target and a technically challenging separation of the $^{99}{\rm Mo}$ produced from the remaining $^{100}{\rm Mo}$~\cite{Hagi18}. As such, $^{96}{\rm Zr}(\alpha,n)$ has been considered as an alternative production route, since the high specific activity of $^{99}{\rm Mo}$ that is produced enables standard $^{99}{\rm Mo}/^{\rm 99m}{\rm Tc}$ generators to be employed~\cite{Pupi15}.

Several previous measurements of the $^{96}{\rm Zr}(\alpha,n)$ cross section have been performed in the energy-range of interest for medical isotope production (i.e. where yields are highest)~\cite{Chow95,Pupi14,Hagi18,Mura19,Vill20,Kiss21}. However, this world data contains several inconsistencies and there has yet to be a consistent physics-based evaluation.

The present work aims to address these concerns. We performed single-target activation cross section measurements of $^{96}{\rm Zr}(\alpha,n)$ from $E_{\alpha}=8-13$~MeV. We performed large-scale Hauser-Feshbach calculations, exploring a large phase-space of statistical model parameters, in order to evaluate our results along with the world data on the $^{96}{\rm Zr}(\alpha,n)$ cross section and the $^{96}{\rm Zr}(\alpha,\alpha)$ differential cross section. We find general agreement between our data, previous measurements, and a statistical description of the $^{96}{\rm Zr}(\alpha,n)$ reaction; however, some discrepancies remain at the lowest reaction energies.

 The paper is structured as follows. We describe our activation measurement in Sec.~\ref{sec::level2}. In Sec.~\ref{sec::level3}, we describe our activation cross section determination and large-scale Hauser-Feshbach calculations. In Sec.~\ref{sec::level4} we present our results, followed by a discussion of the implications for CCSN nucleosynthesis and medical isotope production. We conclude and offer recommendations for future measurements in Sec.~\ref{sec:conclusion}.

\section{\label{sec::level2} Experimental Setup}

 Cross section measurements were performed at the Edwards Accelerator Laboratory at Ohio University~\cite{Meis17} using the activation technique. ${\rm He}^{++}$ ions were produced with an Alphatross ion source, accelerated using the 4.5 MV T-type tandem Pelletron, and impinged on a Zr target enriched to 57.4($\pm$0.2)\% $^{96}{\rm Zr}$ with a zirconium areal density of $n_{\rm t}=6.70$($\pm0.67)\times10^{18}$~atom/cm$^{2}$. The areal density was determined by Rutherford scattering, while the Zr enrichment was quoted by the manufacturer, the National Isotope Development Center, based on inductively coupled plasma mass spectrometry. 
 
 Irradiations were performed at energies between $E_{\alpha}=8-13$~MeV in steps of 1~MeV. Irradiations were performed on the same target with a month or more of cooling time in between, where individual irradiation times lasted between $t_{\rm irr}=3-18$~hr with incident beam currents between $\approx10-150$~nA on-target. Over the course of a single irradiation, the incident beam current was collected with a Faraday cup in the target chamber, measured with a current integrator, and recorded every $\Delta t=30$~s in order to account for the small changes in the incident beam current over time. Prior to each irradiation, the beam was tuned through an empty target frame, such that the full beam current was present on the downstream Faraday cup and no current was detected on the target ladder, ensuring that all incident beam impinged on the Zr target.

\begin{figure}[!htbp]
    \begin{center}
    \includegraphics[height=6.5cm,width=\columnwidth]{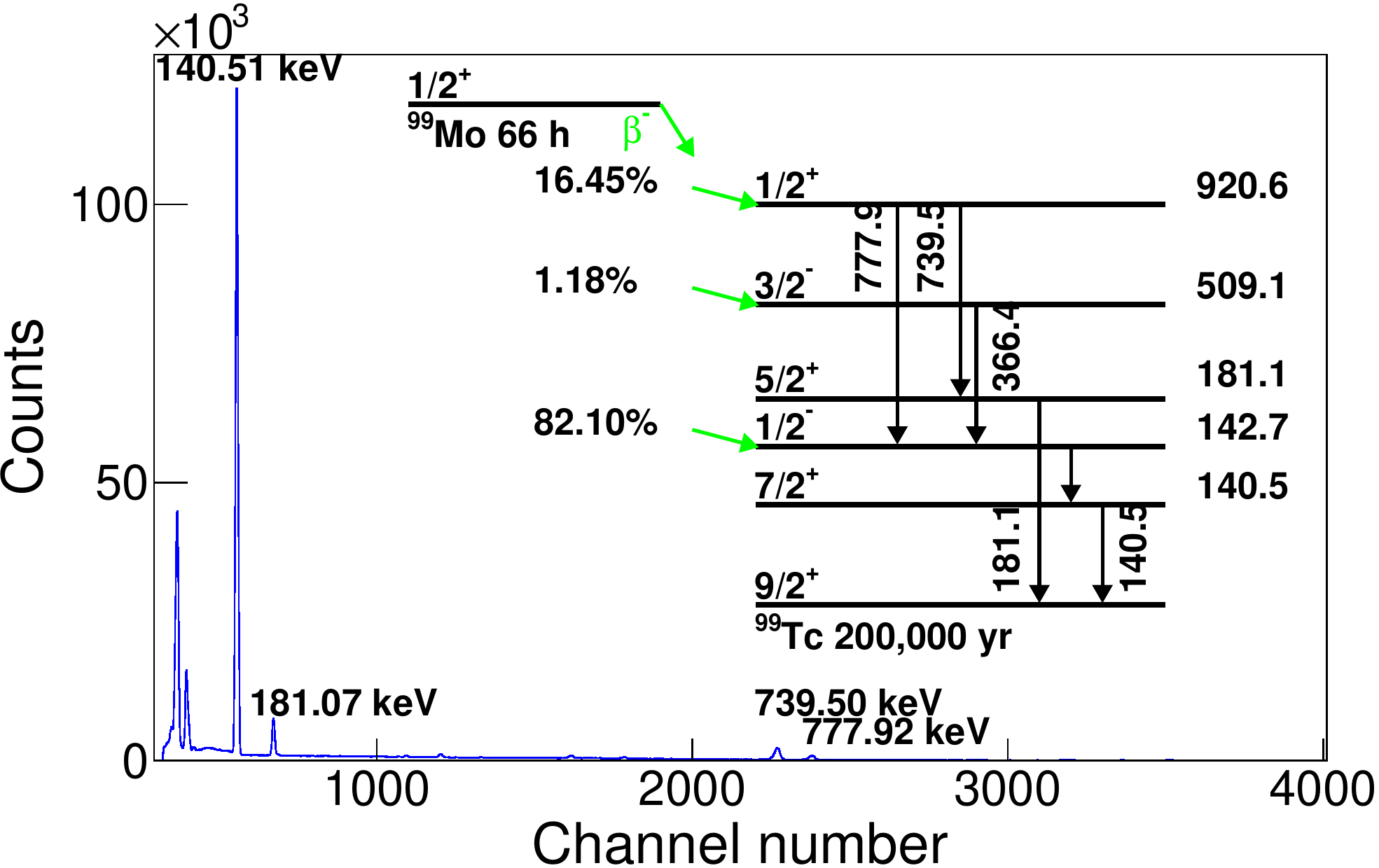}
    \caption{Example $\gamma$-ray spectra from an irradiated Zr target. The peak energies used for the analysis are marked, along with the most intense peak from $\gamma$-decay of $^{99 \rm m}{\rm Tc}$, while the inset shows the $^{99}{\rm Mo}$ decay scheme.}
    \label{fig:decayscheme}
    \end{center}
\end{figure}

After irradiation, the $^{96}{\rm Zr}$ target was transported to a counting station to measure the $\gamma$-activity. The counting station consisted of an 60\% relative efficiency HPGe detector located inside a 4$\pi$ lead shield. The activated target was placed 10~cm from the front face of the detector at $0^{\circ}$. The decay properties for the $^{99}{\rm Mo}$ produced in the activation and its decay product $^{\rm 99m}{\rm Tc}$ are summarized in Table~\ref{table:DecayParam} and Figure~\ref{fig:decayscheme}. We measured the yield of each of these $\gamma$-rays following activation. The $\gamma$-peak area determination is discussed in Section~\ref{sec::level3}.

\begin{table}[h]
 \centering
 \caption{Decay parameters of the reaction product $^{99}{\rm Mo}$ and its daughter $^{\rm 99m}{\rm Tc}$ from Refs.~\cite{Brow17,Gosw92}.}
  \begin{tabular}{@{}cccc@{}}
 \hline
 \hline
    Isotope & Half-life  & Energy & Relative intensity\\
                              & [hr] & [keV] & $\big[ \% \big]$\\
\hline
     $^{99}\rm Mo$& 65.924 $\pm$ 0.006 & 181.07 & 6.05 $\pm$ 0.12\\
&     & 739.50& 1.04 $\pm$ 0.02\\
&     & 777.92 & 4.31 $\pm$ 0.08\\
$^{\rm 99m}\rm Tc$ & 6.0072 $\pm$ 0.0009 & 140.51 & 89 $\pm$ 4\\
\hline
\hline
  \end{tabular}
 \label{table:DecayParam}
\end{table}

\begin{figure}[!htbp]
    \begin{center}
    \includegraphics[width=\columnwidth]{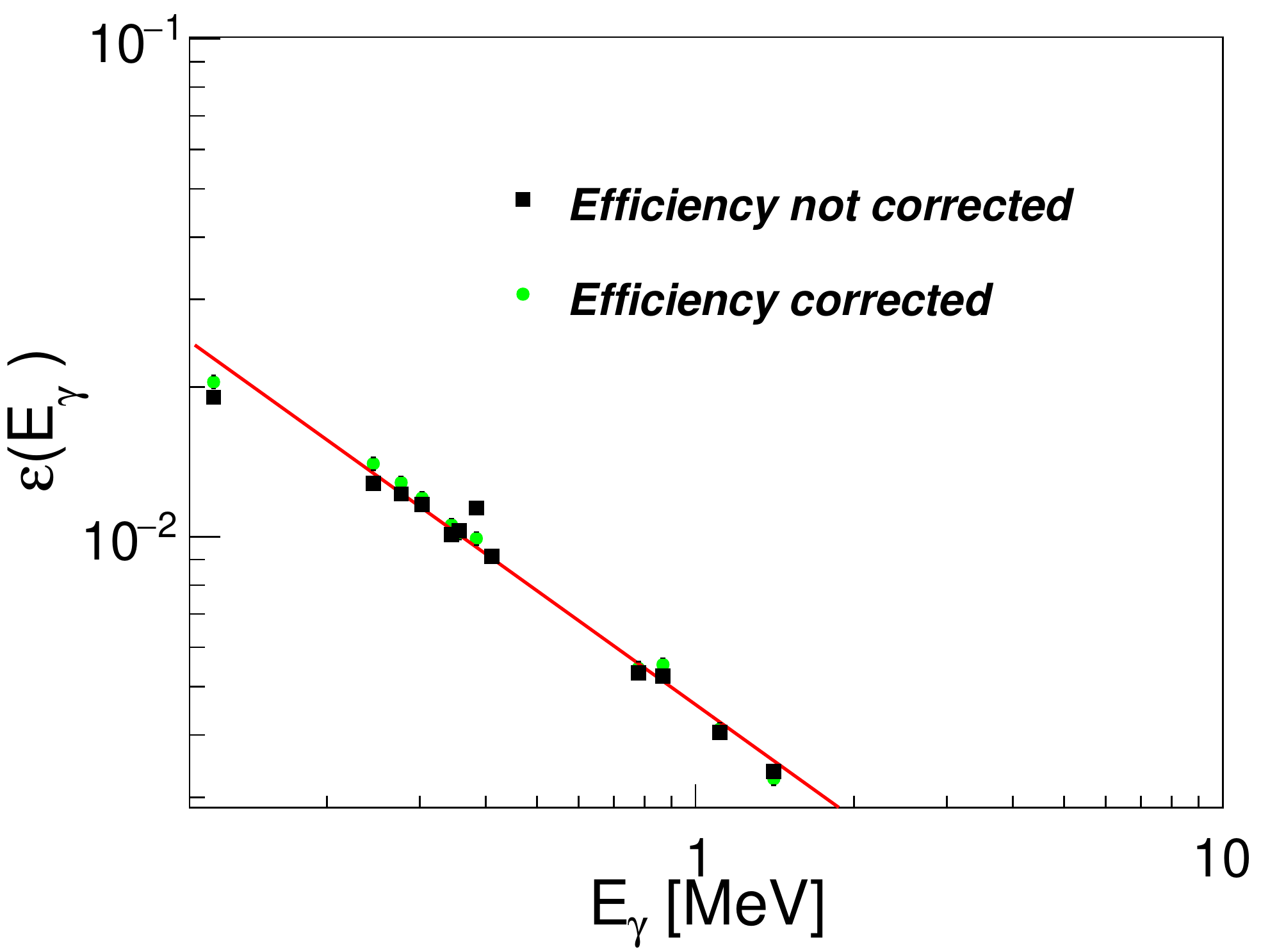}
    \caption{Impact of the coincidence summing correction on $\gamma$-ray photopeak efficiency. The solid red curve is a fit to the corrected data using Equation~\ref{eqn:EffEqn}.}
    \label{fig:summing}
    \end{center}
\end{figure}

The HPGe detector energy and efficiency calibrations were performed using $^{152}\rm Eu$, $^{133}\rm Ba$, $^{60}\rm Co$, and $^{137}\rm Cs$ reference sources from Isotope Products Laboratories, where source activities were certified to $3$\% uncertainty. We removed the impact of coincidence summing on the efficiency using the technique of Ref.~\cite{SEMKOW1990437}. In this technique, the $\beta^{-}$-decay feeding factor, $\gamma$-ray branching factors, and internal-conversion coefficients are supplied for each level based on the known decay scheme. The number of source decays within the counting interval, the number of counts in a selected set of photo-peaks, and the photofraction function $\phi(E_{\gamma})$ for the detector are then provided. The photofraction is the ratio of photopeak efficiency to total efficiency, which we determined with the $^{60}{\rm Co}$ and $^{137}{\rm Cs}$ calibration sources.  
The photopeak efficiencies $\varepsilon(E_{\gamma})$ are then estimated and iterated self-consistently using the coincidence-summing equations until a desired degree of convergence in $\varepsilon(E_{\gamma})$ is reached. The impact of the summing correction is shown in Figure~\ref{fig:summing}.
The empirical function used for the fit of $\gamma$-ray photopeak efficiency is defined as follows,

\begin{equation}\label{eqn:EffEqn}
   \ln(\varepsilon(E_{\gamma}))= a + b \ln( E_{\gamma}) + c\ln(E_{\gamma})^{2}, 
\end{equation}
where $\varepsilon(E_{\gamma})$ is the peak efficiency, $E_{\gamma}$ is $\gamma$-ray energy, and $a$, $b$ and $c$ are fit parameters \cite{knoll}. Here, $a= -5.38$, $b=-3.55$, and $c=-2.04$ with $E_{\gamma}$ in units of MeV.

\section{Analysis}\label{sec::level3}

\subsection{Cross section determination}\label{ssec:Analysisofthespectra}
At each measurement energy, the $^{96}{\rm Zr}(\alpha,n)$ cross section was determined based on the $\alpha$-particle current recorded over the time of the activation and the $\gamma$-ray yield from the activated target following the activation. 

Following the activation measurement, the activity of the target post-irradiation $A_{\rm PO}$ was determined from each $\gamma$-ray of $^{99}{\rm Mo}$ (181.07 keV, 739.50 keV, and 777.92 keV) by evaluating the number of $\gamma$-rays at each of those energies and then averaging the individual results for $A_{\rm PO}$. The exception is our lowest-energy measurement, where we only use the 181.07 keV $\gamma$-ray due to excessive background for the other $\gamma$-rays. We note that results for $A_{\rm PO}$ determined with individual $\gamma$-rays were in agreement within uncertainties, typically differing by $\leq$10\%. For each $\gamma$-ray peak, the number of $\gamma$-rays $N_{\gamma}$ was obtained by fitting the peak with Gaussian and linear functions, subtracting the linear function as the estimated background, integrating the number of remaining counts in the peak region, and accounting for the $\gamma$-detection efficiency, including the effects of coincidence summing. The uncertainty in the number of counts for a given peak is the statistical uncertainty $\sqrt{N_{\gamma}}$ summed with the uncertainty in the $\gamma$-detection efficiency and the systematic uncertainty from the fit:
\begin{equation}
     \sigma_{\rm sys}^{2} = \sum_{i}^{n}  p_{i}^{2}\sigma_{i}^{2} + \sum_{i}^{n}\sum_{j(j\neq i)}^{n} p_{i} p_{j}\rho_{ij}   \sigma_{i}\sigma_{j}.
\end{equation}
Here, $p_{i}$ and $p_{j}$ are the derivatives of the fitting function with respect to its parameters, $\rho_{ij}$ is the correlation between $p_{i}$ and $p_{j}$, and $\sigma_{i}$ and $\sigma_{j}$ are the elements of the covariance matrix. The activity $A$ (Bq) of each radioisotope post-irradiation was calculated according to the following equation:
\begin{equation}
 A_{\rm PO} =\frac{N_\gamma \lambda}{ \varepsilon( E_{\gamma}) I_{\gamma} f_{\rm live}(e^{-\lambda t_{\rm PO}}-e^{-\lambda t_{\rm EOC}})},
\end{equation}
where $\lambda$ is the decay constant (s$^{-1}$), $I_{\gamma}$ is the relative $\gamma$ intensity from Table~\ref{table:DecayParam}, $f_{\rm live}$ is the data acquisition live-fraction determined using a pulser, and $t_{\rm PO}$ and $t_{\rm EOC}$ are the times post-irradiation and of $\gamma$-ray counting, respectively. We note that $t_{\rm EOC}$ is the counting time added to $t_{\rm PO}$.

In principle, we could have determined the $^{96}{\rm Zr}(\alpha,n)$ cross section $\sigma_{\alpha,n}$ using the activation equation for thin targets: $A_{\rm EOI}=I_{\rm b}n_{\rm t}\sigma_{\alpha,n}\left(1-\exp(-\lambda t_{\rm irr})\right)$, where $I_{\rm b}$ is the beam intensity, and the activity at the end of irradiation $A_{\rm EOI}=A_{\rm PO}\exp\left(-\lambda(t_{\rm PO}-t_{\rm EOI})\right)$, with $t_{\rm EOI}$ as the time at the end of irradiation. However, this would not account for variations in $I_{\rm b}$ over the irradiation time, which were generally less than 2\% but occasionally as large as 4\%. Instead, we opted to numerically determine $A_{\rm EOI}$ for a grid of $\sigma_{\alpha,n}$ guesses, which we could then compare to the measured $A_{\rm EOI}\pm\delta A_{\rm EOI}$ in order to determine $\sigma_{\alpha,n}\pm\delta\sigma_{\alpha,n}$. For each $\sigma_{\alpha,n}$ guess, at each time step, the change in the number of $^{99}{\rm Mo}$ nuclei $\Delta N$ was determined by the difference between the production rate $R_{+}$ and destruction rate $R_{-}$. At time $t$, $R_{+}(t)=I_{\rm b}(t)n_{t}\sigma_{\alpha,n}$, using the thin-target approximation, and $R_{-}(t)=\lambda N(t)$, where $\lambda$ is the $^{99}{\rm Mo}$ decay constant and $N(t)$ is the number of $^{99}{\rm Mo}$ at time $t$ since the beginning of the irradiation, with $N(0)=0$. At each time step, $N$ is evolved as:
\begin{equation}
    N(t+1)=N(t)+\Delta N= N(t)+\left(R_{+}(t)-R_{-}(t)\right)\Delta t,
\end{equation}
where $\Delta t$ is the time difference between current readings. This is performed from $t=0$ to $t_{\rm EOI}$, resulting in an estimated $A_{\rm EOI}$. For a single measurement energy, the $\sigma_{\alpha,n}$ that results in agreement between the estimated and measured $A_{\rm EOI}$ is the measured cross section for that energy, while the range of $\sigma_{\alpha,n}$ that are within the measured $\delta A_{\rm EOI}$ provide a portion of the uncertainty $\delta\sigma_{\alpha,n}$. 

The remaining contributions to the cross section uncertainty are due to the uncertainty in the areal density of the target and the uncertainty in the measured current. The total uncertainty was primarily due to the target thickness (10\%), current uncertainty (4\%), and $\gamma$-detection efficiency (3\%). We estimate that the uncertainty contribution from the $\gamma$-summing correction is less than 1\%. To determine the total uncertainty in the cross section, the uncertainty in the target thickness and the efficiency were combined linearly first and then combined in quadrature with the uncertainty in the current. The statistical uncertainty $\sqrt{N_{\gamma}}$ was negligible for all measurements.

\begin{figure}[ht]
    \begin{center}
    \includegraphics[width=\columnwidth]{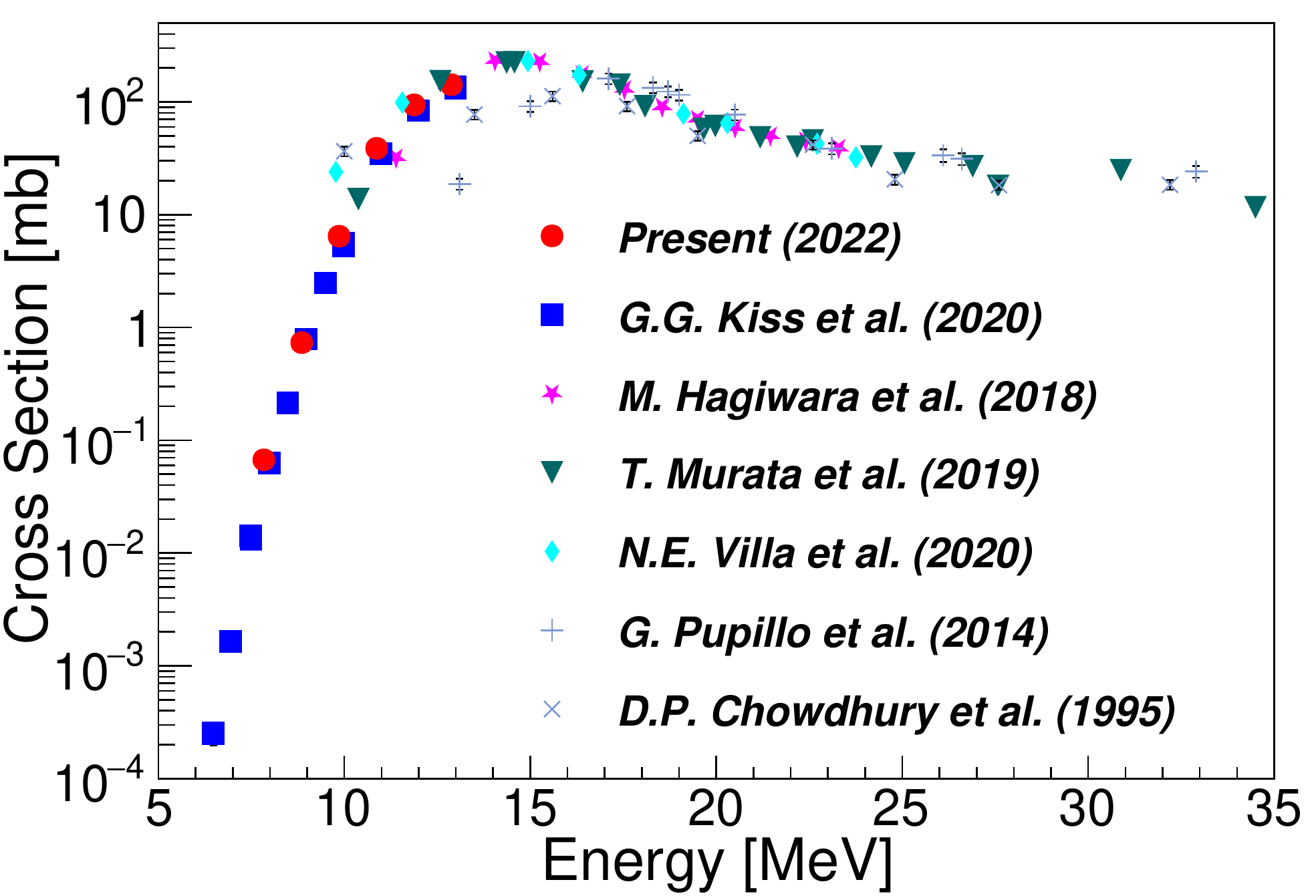}
    \caption{Cross section of $^{96}{\rm Zr}(\alpha,n)$ over a range of $\alpha$ energies in the lab frame as determined in this work (Present) compared to literature values~\cite{Chow95,Pupi14,Hagi18,Mura19,Vill20,Kiss21}.}
    \label{fig:crosssection}
    \end{center}
\end{figure}

\begin{table}[h]
 \centering
 \caption{Cross sections of $^{96}$Zr($\alpha$,n)$^{99}$Mo measured in this work, where the beam on-target energy $E_{\rm beam}$, calculated energy loss of the beam in the target $E_{\rm loss}$, and calculated laboratory energy at the center of the target $E_{\rm lab}$ are in units of MeV.}
   \begin{tabular}{@{}ccccc@{}}
 \hline
 \hline
   E$_{\rm beam}$ & E$_{\rm loss}$  & E$_{\rm lab}$ & &Cross section  [mb]  \\
\hline
7.994 & 0.298 & 7.85$\pm$0.034  & &(6.71 $\pm 0.91$)$\times$10$^{-02}$\\
9.000 & 0.277 & 8.86$\pm$0.034  & &(7.34 $\pm 0.99$)$\times$10$^{-01}$\\
9.997 & 0.259 & 9.87$\pm$0.034  & &(6.45 $\pm 0.88$)$\times$10$^{+00}$\\
10.999 & 0.246 & 10.88$\pm$0.034 & &(3.88 $\pm 0.53$)$\times$10$^{+01}$\\
11.995 & 0.229 & 11.88$\pm$0.034 & &(9.42 $\pm 1.28$)$\times$10$^{+01}$\\
12.996 & 0.220 & 12.89$\pm$0.036 & &(1.42 $\pm 0.19$)$\times$10$^{+02}$\\

\hline
\hline
  \end{tabular}
 \label{table:CSResults}
\end{table}

Our measured cross sections for $^{96}$Zr($\alpha$,n), listed in Table~\ref{table:CSResults}, are compared to results from prior works in Figure~\ref{fig:crosssection}. The reported energy in Table~\ref{table:CSResults} is the lab-frame energy at the center of the target, while the uncertainty reflects fluctuations in the beam energy ($\pm$1~keV), beam energy uncertainty from the opening of the slits following the analyzing magnet ($\pm$0.2\%), and the uncertainty of the energy loss of the beam in the target, including the 10\% uncertainty in the target thickness. We adopt an uncertainty of 4\% for the stopping power of Ref.~\cite{Zieg10} based on the analysis of Ref.~\cite{Paul05} and excellent agreement with the only data set in this energy region~\cite{Lin73}. Our data are largely in agreement with prior works, in particular the recent data of Ref.~\cite{Kiss21}. However, we diverge somewhat from those results at our lowest measurement energy. We discuss this further in Section~\ref{sec::level4}.

\subsection{Hauser-Feshbach calculations}

In order to achieve a physics-based evaluation of the data in Figure~\ref{fig:crosssection}, we performed large-scale Hauser-Feshbach calculations~\cite{Hauser1,Wolf51} with the code {\tt Talys} v1.95~\cite{Koni08}. The goal of these calculations was to find a consistent description of the $^{96}{\rm Zr}(\alpha,n)$ world data, while simultaneously consistently describing differential cross section data from $^{96}{\rm Zr}(\alpha,\alpha)$~\cite{Laha86,Lund95}, as the latter data is similarly sensitive to Hauser-Feshbach input parameters. For a comparison between the $^{96}{\rm Zr}(\alpha,n)$ cross section world data and results calculated using standard global $\alpha$-optical potentials, see Ref.~\cite{Kiss21}.

In the Hauser-Feshbach formalism, the $(\alpha,n)$ cross section is described by $\sigma_{\rm \alpha,n}\propto\lambda_{\alpha}^{2}(\mathcal{T}_{\rm \alpha}\mathcal{T}_{n})/\Sigma_{j }\mathcal{T}_{{\rm decay,}j}$, where $\lambda_{\alpha}$ is the de Broglie wavelength for the incident $\alpha$ and the $\mathcal{T}_{i}$ are the transmission coefficients that describe the probability for a particle or photon, which defines the channel, being emitted from (``decay") or absorbed by a nucleus. The $\mathcal{T}$ for decay channels in the preceding equation,  including the neutron transmission coefficient $\mathcal{T}_{n}$,  and all other open decay channels included in the sum $\Sigma_{j}\mathcal{T}_{{\rm decay},j}$, are actually a sum over $\mathcal{T}$ to individual discrete states added to an integral over $\mathcal{T}$ to levels in a higher-excitation energy region described by the nuclear level-density $\rho$ (See e.g. Equation~3 of Ref.~\cite{Lars19}). For the energies of interest in this work, $\Sigma_{j}\mathcal{T}_{{\rm decay},j} \approx \mathcal{T}_{n}$ and most other ingredients of the Hauser-Feshbach input are expected to play a minor role~\cite{Mohr16}. As such, we primarily focused on varying the parameters of the $\alpha$-optical potential, which is used to calculate the $\alpha$ transmission coefficient $\mathcal{T}_{\alpha}$. We also explored modifications to $\rho$ and to the level-spin distribution, via the spin-cutoff parameter $\sigma_{\rm sc}^{2}$, of $^{99}{\rm Mo}$, as these can impact the competition between $^{96}{\rm Zr}(\alpha,n)^{99}{\rm Mo}$ and $^{96}{\rm Zr}(\alpha,2n)^{98}{\rm Mo}$ channels for $E_{\alpha}\gtrsim13$~MeV.

We adopted the McFadden and Satchler~\cite{mcfadden} parameterization of the $\alpha$-optical potential for our study, as the simple functional form is preferable given our limited data set and the fact that the global parameterization of that work generally provides a good description of $(\alpha,n)$ reaction cross sections~\cite{Mohr15}. In this description, the $\alpha$-optical potential is 
\begin{equation}
    U(r)=V_{\rm c}(r)+V(r)+iW(r),
\end{equation}
where $V_{\rm c}(r)$ is the Coulomb potential and $V(r)$ and $W(r)$ are the real and imaginary parts of the nuclear potential, respectively. The latter two are described with a Woods-Saxon form,
\begin{equation}
    \begin{split}
    V(r)&=\frac{-V}{1+\exp\left((r-r_{v})/a_{v}\right)}, \\
    W(r)&=\frac{-W}{1+\exp\left((r-r_{w})/a_{w}\right)},
\end{split}
\end{equation}
where $V$ and $W$ describe the potential well depths and $r_{i}$ and $a_{i}$ describe the potential radius and diffuseness, respectively. Following the approach of Ref.~\cite{mcfadden}, we set $r_{w}=r_{v}$ and $a_{w}=a_{v}$.

For $\rho$, we adopted the back-shifted Fermi gas (BSFG) model~\cite{Dilg73}, as this option most closely reproduced the data of Figure~\ref{fig:crosssection} when using {\tt Talys} default parameters otherwise. Additionally, the BSFG model appears to adequately reproduce $\rho$ for nuclei in this region of the nuclear chart~\cite[e.g.][]{Chank06,Mart17}. For the BSFG model, 
\begin{equation}
\rho(E^{*})=\frac{\exp{\left(2\sqrt{a(E^{*}-\Delta_{\rm bs})}\right)}}{12\sqrt{2\sigma_{\rm sc}^{2}} a^{1/4}(E^{*}-\Delta_{\rm bs})^{5/4}},
\label{eqn:rhobsfg}
\end{equation}
where $\Delta_{\rm bs}$ is a back-shift to ensure $\rho$ is described at low-lying excitation energies where all discrete levels are known and $a$ is the excitation energy ($E^{*}$)-dependent level density parameter of Ref.~\cite{Igna75}, based on global fits to $\rho$. In {\tt Talys}, one can set $a$ at the neutron separation energy $S_{n}$ and then the $a$ at other $E^{*}$ will scale accordingly.  The spin distribution is the rigid body form~\cite{Beth36},
\begin{equation}
    P(J)=\frac{2J+1}{2\sigma_{\rm sc}^{2}}\exp\left(\frac{-(J+1/2)^{2}}{2\sigma_{\rm sc}^{2}}\right),
\end{equation}
which is used to convert from $\rho$ to a density of levels with spin $J$, where $\rho(E^{*},J)=P(J)\rho(E^{*})$. The $E^{*}$-dependent $\sigma_{\rm sc}^{2}$ is determined in {\tt Talys} by three different methods, depending on the $E^{*}$ region. For low $E^{*}$, where all levels are thought to be known (here $E^{*}\leq1.5$~MeV~\cite{Capo09}), 
 the discrete level region  $\sigma_{\rm sc}^{2}$ is used. Here,
\begin{equation}
    \sigma_{\rm sc,disc}^{2}=\frac{\sum{ J_{i}(J_{i}+1)(2J_{i}+1)}}{3\sum{2J_{i}+1}},
    \label{eqn:SigDisc}
\end{equation}
where the sum runs over all levels in the discrete level region. For $E^{*}\geq S_{n}$, the Fermi gas estimate is used,
\begin{equation}
    \sigma_{\rm sc,FG}^{2}\approx0.04A^{7/6}\sqrt{E^{*}},
    \label{eqn:SigFG}
\end{equation}
where $A$ is the mass number. The exact form for Equation~\ref{eqn:SigFG} is available in the {\tt Talys} manual. For intermediate $E^{*}$, $\sigma_{\rm sc}^{2}$ is a linear interpolation between $\sigma_{\rm sc,disc}^{2}$ and $\sigma_{\rm sc,FG}^{2}$.

\begin{table}[h]
\centering
\caption{Hauser-Feshbach parameters varied within a randomly-sampled uniform range, along side the nominal values for comparison. The values resulting in the minimum global $\chi^{2}$ are also provided.}
\begin{tabular}{@{}ccccc@{}}
\hline
\hline
Parameter &  Nominal  &  Range & Best Fit  & Best Fit \\
          &           &        &  Including  & Not Including  \\
           &           &        &   Ref.~\cite{Kiss21}&  Ref.~\cite{Kiss21} \\

\hline
 $s2$  &1 & 1 -- 1.9& 1.41  & 1.33 \\
 $a(S_{n})$  & 12.43 &  11.37 -- 12.4& 11.34  &11.34  \\
 $V$ &  185 &  140 -- 220 &  181.67 & 195.78\\
 $W$ &  25 &  5 -- 54 & 17.51  & 17.88 \\
$r_{v}$ &  1.4 &  1.2 -- 1.6 & 1.43  & 1.32\\
$a_{v}$ &  0.52 &  0.4 -- 0.7 & 0.48 & 0.59 \\
\hline
\hline
\end{tabular}
\label{table:AOPparameters}
\end{table}

\begin{figure*}[htbp]
\begin{center}
\includegraphics[width=0.95\textwidth]{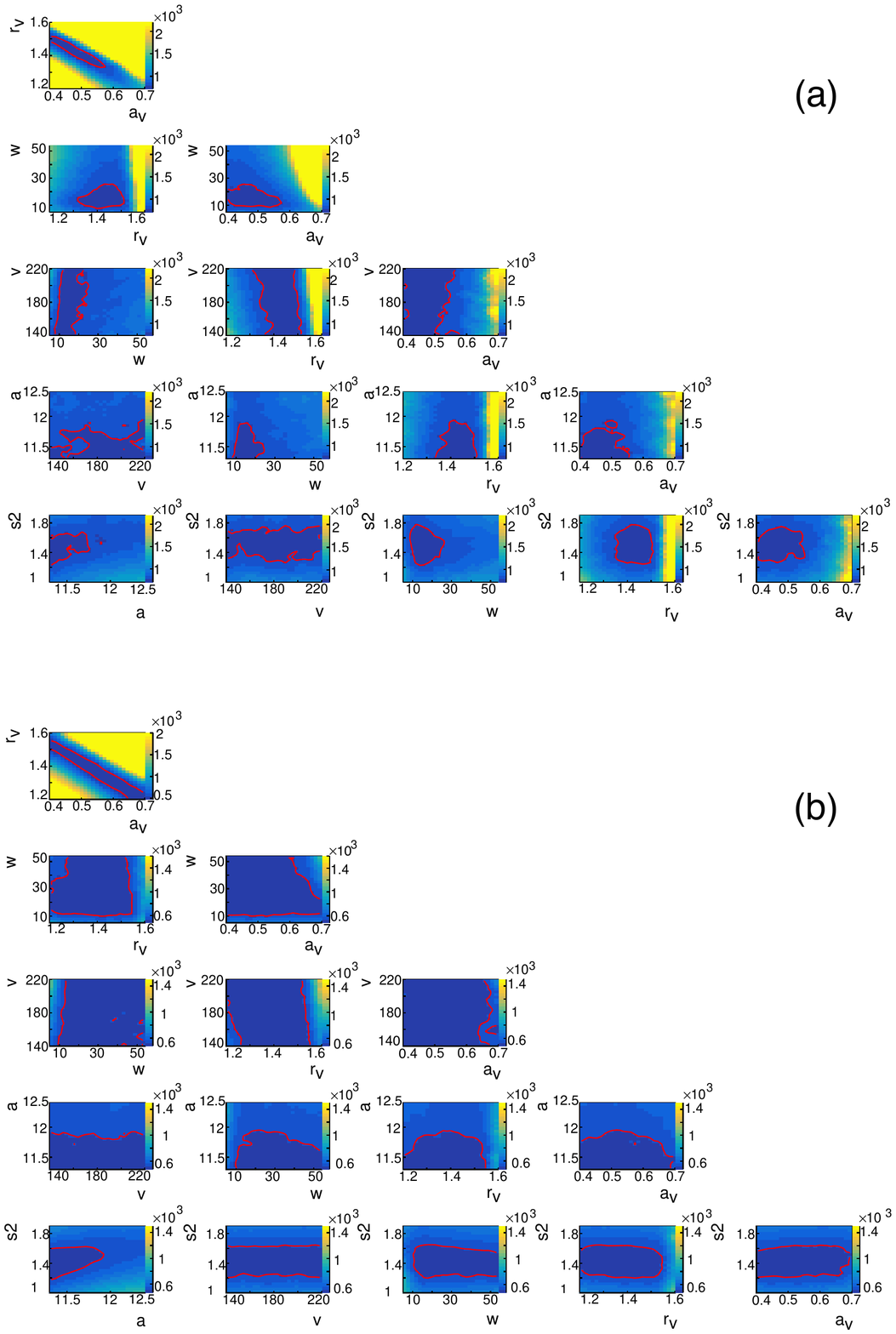}
 \caption{ (a) Corner plot of $\chi^{2}$ calculated when comparing results of our {\tt Talys} calculations to the $^{96}{\rm Zr}(\alpha,n)$ cross section data of Figure~\ref{fig:crosssection}, excluding the data of Ref.~\cite{Chow95} and Ref.~\cite{Pupi14}, and projected into the two-parameter phase space of each panel. For each panel, the parameters that are not indicated on the axes of that panel have been randomly sampled within the range described by Table~\ref{table:AOPparameters}. The $\chi^{2}$ shown in each bin of each histogram is the minimum $\chi^{2}$ of each of the $100\,000$ Hauser-Feshbach calculations that employ parameters within that bin. The red contours indicate the 95\% confidence intervals, as calculated by $\Delta\chi^{2}$. (b) Same as sub-figure (a), but also excluding the data of Ref.~\cite{Kiss21} in the $\chi^{2}$ determinations. }
 \label{fig:chi2_CS}
\end{center}
\end{figure*}

 \begin{figure*}[htbp]
\begin{center}
\includegraphics[width=0.95\textwidth]{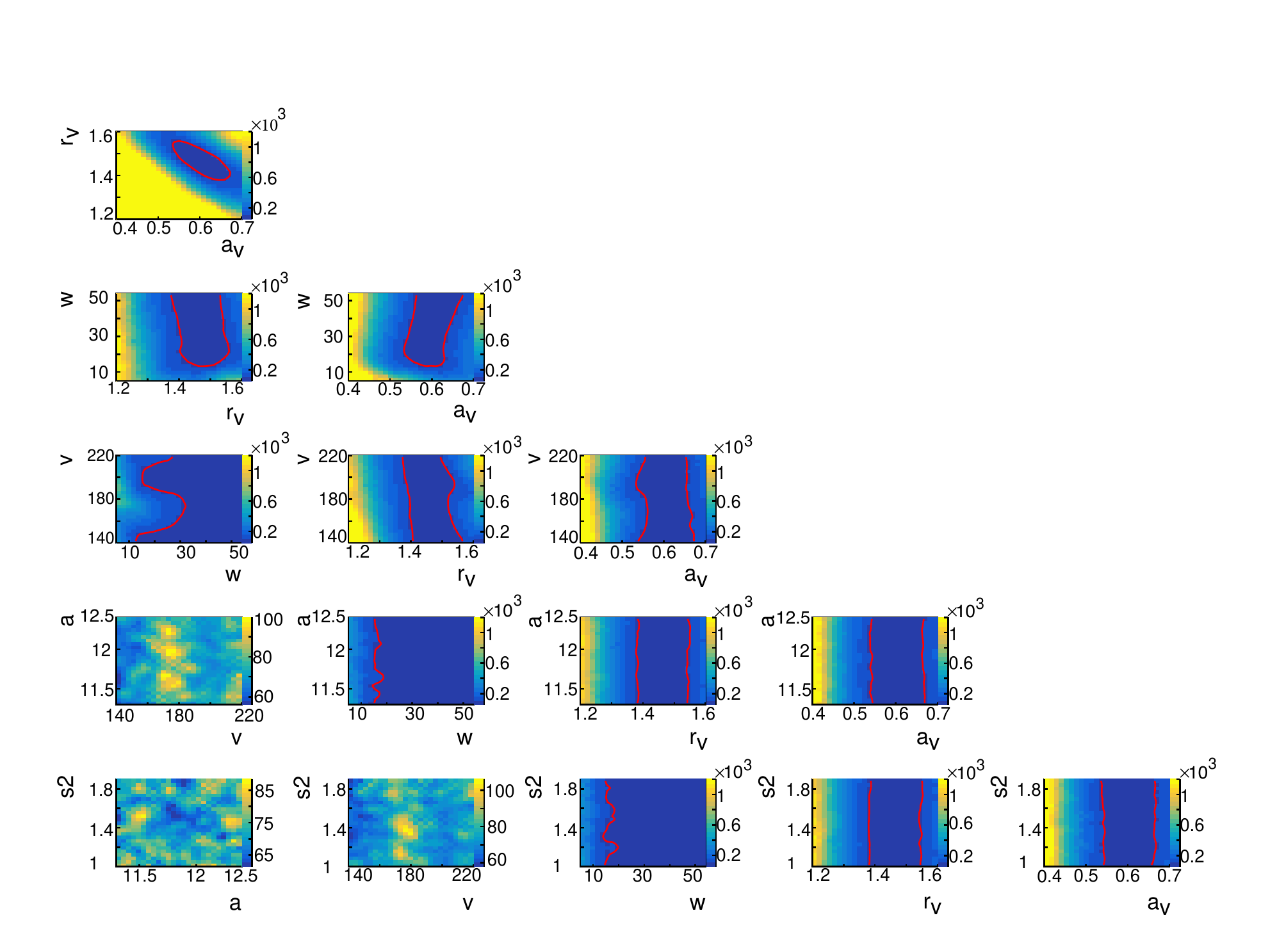}
 \caption{Corner plot of $\chi^{2}$ calculated when comparing results of our {\tt Talys} calculations to the $^{96}{\rm Zr}(\alpha,\alpha)$ differential cross section data of Figure~\ref{fig:diffCS} and projected into the two-parameter phase space of each panel. For each panel, the parameters that are not indicated on the axes of that panel have been randomly sampled within the range described by Table~\ref{table:AOPparameters}. The $\chi^{2}$ shown in each bin of each histogram is the minimum $\chi^{2}$ of each of the $100\,000$ Hauser-Feshbach calculations that employ parameters within that bin. The red contours indicate the 95\% confidence intervals, as calculated by $\Delta\chi^{2}$. When red contours are absent in a panel, all parameter combinations are within the 95\% confidence interval.}
 \label{fig:chi2_diffCS}
\end{center}
\end{figure*}

In an attempt to simultaneously describe existing total cross section data for $^{96}{\rm Zr}(\alpha,n)$ and differential cross section data for $^{96}{\rm Zr}(\alpha,\alpha)$, we varied $V$, $W$, $r_{v}$, $a_{v}$, $a(S_{n})$, and $\sigma^{2}_{\rm sc}(S_{n})$. We performed $100\,000$ calculations with {\tt Talys}, where each of these parameters was sampled from a uniform distribution within a range stipulated in Table~\ref{table:AOPparameters}. In that table, $s2$ is a multiplier of $\sigma_{\rm sc}^{2}(S_{n})$ from Equation~\ref{eqn:SigFG}. For $V$, $W$, $r_{v}$, and $a_{v}$, the centroids of our randomly sampled ranges roughly correspond to the nominal values from Ref.~\cite{mcfadden} (listed in Table~\ref{table:AOPparameters}). The upper and lower bounds are based on a systematic investigation of the parameter space. In this investigation, we originally chose some parameter ranges, performed $100\,000$ Hauser-Feshbach calculations, computed the chi-square $\chi^{2}$ between the calculation results and the data, and then expanded the parameter range and repeated our calculations until we saw a divergence in $\chi^{2}$ near the boundary of the parameter space. For $a(S_{n})$ our range was guided by the level-density trends for Mo isotopes reported by Ref.~\cite{Chank06}. While systematics in data and theory justify a range for $s2$ between 0.5--2~\cite{Grim16}, we found that only an increase in $\sigma_{\rm sc}^{2}(S_{n})$ improved agreement with the $^{96}{\rm Zr}(\alpha,n)$ cross section data (from the nominal values) and so we did not explore $s2<1$ in our final set of calculations.

For each of the $100\,000$ Hauser-Feshbach calculations, we computed separate $\chi^{2}$ for the $^{96}{\rm Zr}(\alpha,n)$ cross section and for the $^{96}{\rm Zr}(\alpha,\alpha)$ differential cross section. For the latter, we used the data of Refs.~\cite{Laha86,Lund95}, each of which were obtained for $E_{\alpha}=35.4$~MeV. For the former, we included all of the data shown in Figure~\ref{fig:crosssection}, except for the data of Ref.~\cite{Chow95} due to its large deviation from other data sets. Partly motivated by the discrepancy between our low $E_{\alpha}$ results and those of Ref~\cite{Kiss21}, we also performed $\chi^{2}$ calculations when additionally excluding the data from Ref.~\cite{Kiss21}. In order to obtain constraints for the parameters in Table~\ref{table:AOPparameters}, we computed confidence intervals using $\Delta\chi^{2}=\chi^{2}-\chi^{2}_{\rm min}$, where $\chi^{2}_{\rm min}$ is the minimum $\chi^{2}$ across the explored phase-space and the mapping between $\Delta\chi^{2}$ and a confidence interval depends on the number of degrees of freedom~\cite{Pres92}.

Figures~\ref{fig:chi2_CS} and \ref{fig:chi2_diffCS}  show the results of our $\chi^{2}$ calculations. In these figures, the minimum $\chi^{2}$ of the calculations that fall within each bin of each histogram are used to produce the color contours. The red-line boundaries indicate the 95\% confidence interval as determined by $\Delta\chi^{2}$. Figure~\ref{fig:chi2_CS} sub-figures (a) and (b)  are for the $^{96}{\rm Zr}(\alpha,n)$ cross section data including and excluding the data from Ref.~\cite{Kiss21}, respectively. Figure~\ref{fig:chi2_diffCS} is for the $^{96}{\rm Zr}(\alpha,\alpha)$ differential cross section data.   
As discussed further in Section~\ref{sec::level4}, the 95\% confidence intervals determined from comparison to $^{96}{\rm Zr}(\alpha,n)$ cross section data and $^{96}{\rm Zr}(\alpha,\alpha)$ differential cross section data do not overlap for some parameters. For instance, compare the $r_{v}$ versus $a_{v}$ panel of Figure~\ref{fig:chi2_CS} (a) or (b) to the same panel of Figure~\ref{fig:chi2_diffCS}. As the focus of our work is a determination of the $^{96}{\rm Zr}(\alpha,n)$ cross section and astrophysical reaction rate, we decided to focus on parameter combinations within the 95\% confidence intervals of Figure~\ref{fig:chi2_CS} when evaluating a recommended cross section and reaction rate.

\begin{figure*}[]
\begin{center}
\includegraphics[width=2\columnwidth]{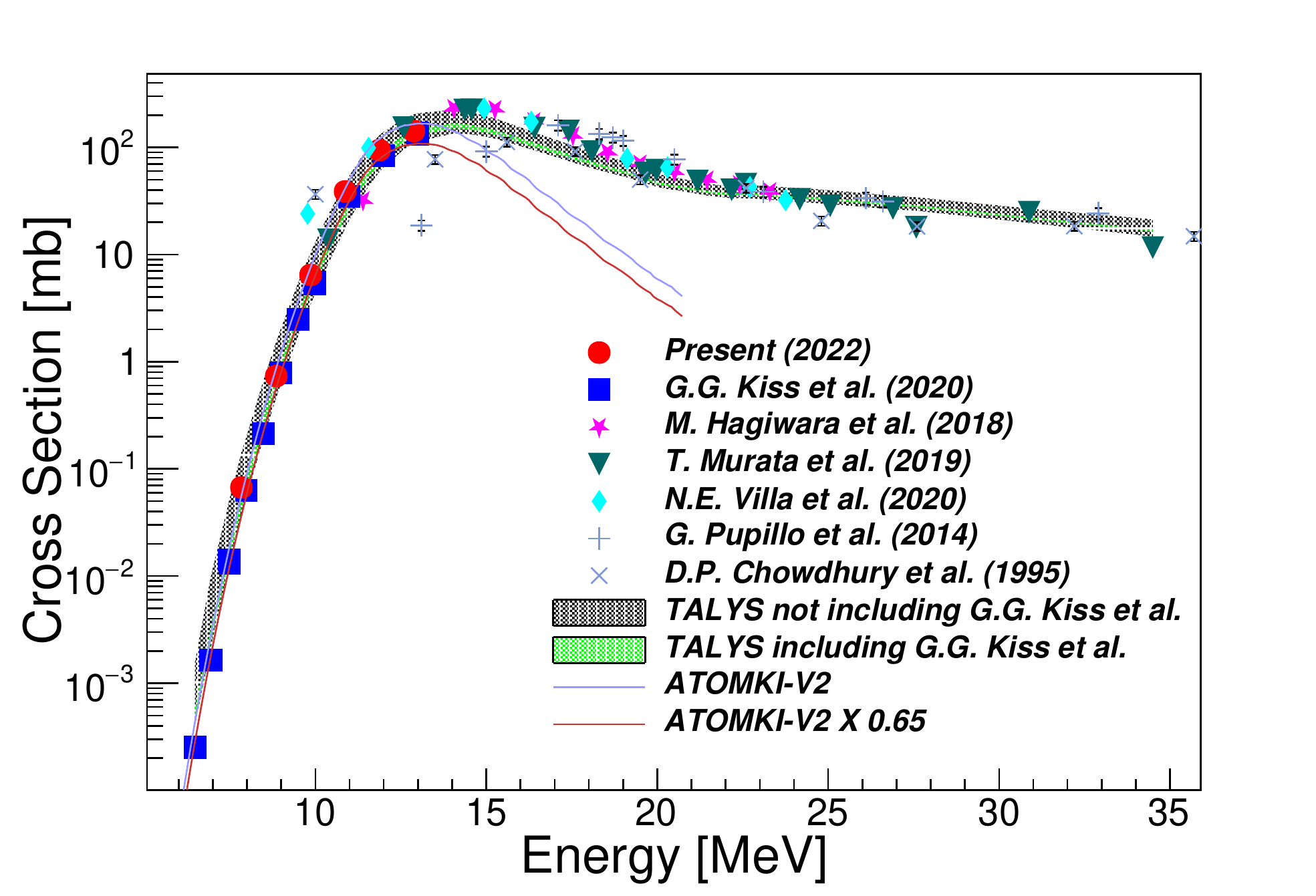}
\caption{Same as Figure~\ref{fig:crosssection}, with the addition of Hauser-Feshbach calculation results from this work and the calculations of Ref.~\cite{Kiss21} that employed the ATOMKI-V2 potential (red line). The green (black) band includes (excludes) the measurement results of Ref.~\cite{Kiss21} in the $\chi^{2}$ calculations used to determine the best-fit parameters sampled from Table~\ref{table:AOPparameters}.}
\label{fig:CSbands}
\end{center}
\end{figure*}

\begin{figure*}[]
\begin{center}
\includegraphics[width=2\columnwidth]{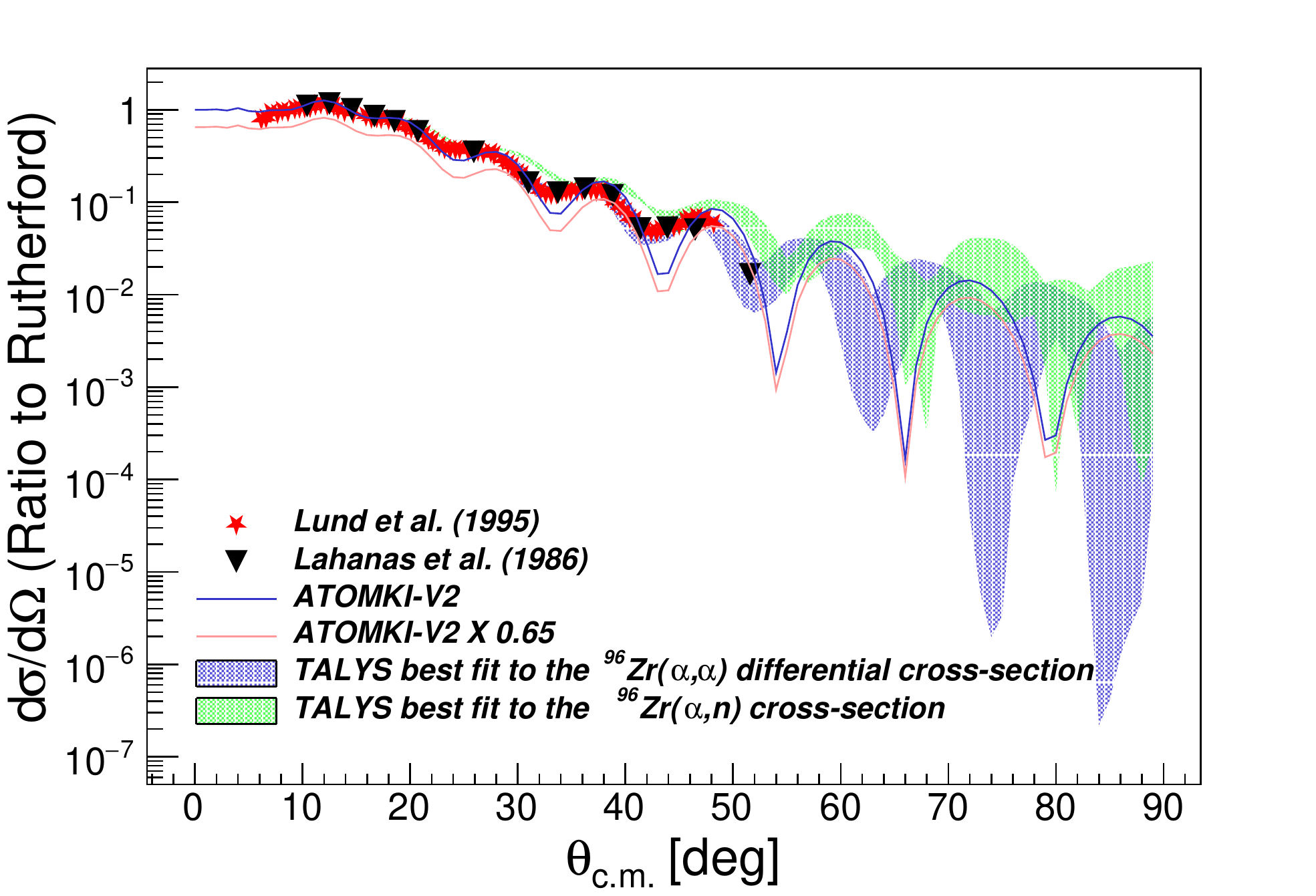}
\caption{Elastic scattering cross section $^{96}{\rm Zr}(\alpha,\alpha)$ at $E_{\alpha}=35.4$~MeV from Ref.~\cite{Laha86} (Lahanas) and Ref.~\cite{Lund95} (Lund), compared to {\tt Talys} calculation results from this work. The green band corresponds to results from calculations whose parameters fall within all 95\% confidence interval contours of Figure~\ref{fig:chi2_CS}. The lavender band corresponds to results from calculations whose parameters fall within all 95\% confidence interval contours of Figure~\ref{fig:chi2_diffCS}. The red line was calculated using the ATOMKI-V2 potential.}
\label{fig:diffCS}
\end{center}
\end{figure*}

We determined an evaluated $^{96}{\rm Zr}(\alpha,n)$ cross section by creating an uncertainty band based on all Monte Carlo iterations that fell within all of the 95\% confidence interval contours of Figure~\ref{fig:chi2_CS}a. We also determined a rate uncertainty band when only considering the 95\% confidence interval contours of Figure~\ref{fig:chi2_CS}b. 
The resulting cross section uncertainty bands are compared to the $^{96}{\rm Zr}(\alpha,n)$ cross section world data in Figure~\ref{fig:CSbands}. In Figure~\ref{fig:diffCS} we compare the $^{96}{\rm Zr}(\alpha,\alpha)$ differential cross section world data to two calculations results: (1) an uncertainty band based on all Monte Carlo iterations that fall within the 95\% confidence interval contours of Figure~\ref{fig:chi2_CS}b, and (2) an uncertainty band based on all Monte Carlo iterations that fall within the 95\% confidence interval contours of Figure~\ref{fig:chi2_diffCS}.

For our best-fit to the $^{96}{\rm Zr}(\alpha,n)$ cross section data when including Ref.~\cite{Kiss21}, the chi-square per point $\chi^{2}_{N}=14.69$. When excluding Ref.~\cite{Kiss21}, the best-fit resulted in $\chi^{2}_{N}=7.86$. This indicates a relatively poor goodness of fit, which is in part due to the large scatter of experimental data shown in Fig.~\ref{fig:crosssection}. To characterize the goodness-of-fit for the high-precision data in the energy region of astrophysical interest, can also calculate this statistic when including only the data from this work and Ref.~\cite{Kiss21}. The calculated cross section resulting from the best-fit that includes Ref.~\cite{Kiss21} in the fit is $\chi^{2}_{N}=5.63$, while the best-fit cross section that omitted Ref.~\cite{Kiss21} in the fit is $\chi^{2}_{N}=1.31$. The corresponding Hauser-Feshbach model parameters are listed in Table~\ref{table:AOPparameters}. The evaluated cross section results for the best-fits, along with the upper and lower bounds of the 95\% confidence interval contours, are reported in Table~\ref{tab:EvaluatedCSResults}.

\begin{table}[htpb]
 \centering
 \caption{Evaluated cross-section for $^{96}$Zr($\alpha$,n)$^{99}$Mo in units of mb and $\alpha$ laboratory energy in MeV. The best-fit and upper(lower) cross-section are reported for the evaluated results that include ($\sigma$ A) or exclude ($\sigma$ B) Ref.~\cite{Kiss21}, along with uncertainties (unc.).}\label{tab:EvaluatedCSResults}
   \begin{tabular}{@{}ccccc@{}}
 \hline
 \hline
      $E_{\alpha}$  &  $\sigma$ A & unc. A &  $\sigma$ B &  unc. B \\
\hline
6.94$\times$10$^{+00}$   &    2.78$\times$10$^{-03}$    &    4.46$\times$10$^{-04}$   &    5.94$\times$10$^{-03}$    &    3.82$\times$10$^{-03}$    \\
6.95$\times$10$^{+00}$    &    2.89$\times$10$^{-03}$    &    4.63$\times$10$^{-04}$    &    6.15$\times$10$^{-03}$    &    3.95$\times$10$^{-03}$    \\
7.48$\times$10$^{+00}$    &    1.69$\times$10$^{-02}$    &    2.60$\times$10$^{-03}$    &    3.20$\times$10$^{-02}$    &    1.92$\times$10$^{-02}$    \\
7.49$\times$10$^{+00}$   &    1.75$\times$10$^{-02}$    &    2.68$\times$10$^{-03}$    &    3.30$\times$10$^{-02}$    &    1.97$\times$10$^{-02}$    \\
7.84$\times$10$^{+00}$    &    5.01$\times$10$^{-02}$    &    7.52$\times$10$^{-03}$    &    9.96$\times$10$^{-02}$    &    5.32$\times$10$^{-02}$    \\
7.98$\times$10$^{+00}$   &    7.33$\times$10$^{-02}$   &    1.10$\times$10$^{-02}$    &    1.26$\times$10$^{-01}$    &    7.26$\times$10$^{-02}$    \\
8.48$\times$10$^{+00}$    &    2.69$\times$10$^{-01}$    &    3.98$\times$10$^{-02}$   &    4.30$\times$10$^{-01}$    &    2.39$\times$10$^{-01}$    \\
8.85$\times$10$^{+00}$    &    6.46$\times$10$^{-01}$    &    9.40$\times$10$^{-02}$    &    9.87$\times$10$^{-01}$    &    5.32$\times$10$^{-01}$    \\
8.98$\times$10$^{+00}$    &    8.61$\times$10$^{-01}$    &    1.24$\times$10$^{-01}$    &    1.30$\times$10$^{+00}$    &    6.93$\times$10$^{-01}$    \\
9.49$\times$10$^{+00}$   &    2.48$\times$10$^{+00}$    &    3.49$\times$10$^{-01}$    &    3.56$\times$10$^{+00}$    &    1.84$\times$10$^{+00}$    \\
9.77$\times$10$^{+00}$    &    4.26$\times$10$^{+00}$    &    5.85$\times$10$^{-01}$    &    5.94$\times$10$^{+00}$    &    3.02$\times$10$^{+00}$    \\
9.86$\times$10$^{+00}$    &    5.04$\times$10$^{+00}$    &    6.87$\times$10$^{-01}$    &    6.98$\times$10$^{+00}$    &    3.53$\times$10$^{+00}$    \\
9.98$\times$10$^{+00}$    &    6.16$\times$10$^{+00}$    &    8.31$\times$10$^{-01}$    &    8.44$\times$10$^{+00}$    &    4.23$\times$10$^{+00}$    \\
9.99$\times$10$^{+00}$    &    6.27$\times$10$^{+00}$    &    8.45$\times$10$^{-01}$    &    8.59$\times$10$^{+00}$    &    4.30$\times$10$^{+00}$    \\
1.04$\times$10$^{+01}$    &    1.20$\times$10$^{+01}$   &    1.55$\times$10$^{+00}$   &    1.60$\times$10$^{+01}$    &    7.72$\times$10$^{+00}$    \\
1.09$\times$10$^{+01}$   &    2.50$\times$10$^{+01}$    &    3.01$\times$10$^{+00}$    &    3.21$\times$10$^{+01}$    &    1.46$\times$10$^{+01}$    \\
1.10$\times$10$^{+01}$    &    2.93$\times$10$^{+01}$   &    3.46$\times$10$^{+00}$    &    3.74$\times$10$^{+01}$    &    1.67$\times$10$^{+01}$    \\
1.14$\times$10$^{+01}$    &    4.86$\times$10$^{+01}$    &    5.28$\times$10$^{+00}$   &    6.05$\times$10$^{+01}$    &    2.50$\times$10$^{+01}$    \\
1.16$\times$10$^{+01}$    &    5.85$\times$10$^{+01}$    &    6.11$\times$10$^{+00}$    &    7.20$\times$10$^{+01}$    &    2.88$\times$10$^{+01}$    \\
1.19$\times$10$^{+01}$    &    7.85$\times$10$^{+01}$    &    7.55$\times$10$^{+00}$    &    9.49$\times$10$^{+01}$    &    3.53$\times$10$^{+01}$    \\
1.20$\times$10$^{+01}$    &    8.47$\times$10$^{+01}$    &    7.94$\times$10$^{+00}$    &    1.02$\times$10$^{+02}$    &    3.71$\times$10$^{+01}$    \\
1.26$\times$10$^{+01}$    &    1.20$\times$10$^{+02}$    &    9.64$\times$10$^{+00}$    &    1.41$\times$10$^{+02}$    &    4.50$\times$10$^{+01}$    \\
1.28$\times$10$^{+01}$    &    1.30$\times$10$^{+02}$    &    9.90$\times$10$^{+00}$    &    1.51$\times$10$^{+02}$    &    4.65$\times$10$^{+01}$    \\
1.29$\times$10$^{+01}$    &    1.34$\times$10$^{+02}$    &    9.98$\times$10$^{+00}$    &    1.56$\times$10$^{+02}$   &    4.69$\times$10$^{+01}$    \\
1.30$\times$10$^{+01}$   &    1.39$\times$10$^{+02}$    &    1.00$\times$10$^{+01}$    &    1.61$\times$10$^{+02}$    &    4.74$\times$10$^{+01}$    \\
1.41$\times$10$^{+01}$    &    1.59$\times$10$^{+02}$    &    9.06$\times$10$^{+00}$    &    1.80$\times$10$^{+02}$    &    4.49$\times$10$^{+01}$    \\
1.44$\times$10$^{+01}$    &    1.57$\times$10$^{+02}$    &    8.27$\times$10$^{+00}$    &    1.76$\times$10$^{+02}$    &    4.18$\times$10$^{+01}$    \\
1.46$\times$10$^{+01}$    &    1.53$\times$10$^{+02}$    &    7.88$\times$10$^{+00}$    &    1.71$\times$10$^{+02}$    &    3.97$\times$10$^{+01}$    \\
1.49$\times$10$^{+01}$    &    1.46$\times$10$^{+02}$    &    7.25$\times$10$^{+00}$    &    1.62$\times$10$^{+02}$    &    3.59$\times$10$^{+01}$    \\
1.53$\times$10$^{+01}$    &    1.38$\times$10$^{+02}$    &    6.74$\times$10$^{+00}$   &    1.53$\times$10$^{+02}$    &    3.27$\times$10$^{+01}$    \\
1.63$\times$10$^{+01}$   &    1.08$\times$10$^{+02}$   &    4.99$\times$10$^{+00}$    &    1.19$\times$10$^{+02}$    &    2.29$\times$10$^{+01}$   \\
1.64$\times$10$^{+01}$    &    1.06$\times$10$^{+02}$    &    4.86$\times$10$^{+00}$    &    1.16$\times$10$^{+02}$    &    2.23$\times$10$^{+01}$    \\
1.64$\times$10$^{+01}$    &    1.05$\times$10$^{+02}$   &    4.83$\times$10$^{+00}$   &    1.16$\times$10$^{+02}$    &    2.22$\times$10$^{+01}$    \\
1.74$\times$10$^{+01}$   &    8.32$\times$10$^{+01}$   &    3.44$\times$10$^{+00}$    &    9.08$\times$10$^{+01}$    &    1.62$\times$10$^{+01}$    \\
1.75$\times$10$^{+01}$    &    8.06$\times$10$^{+01}$    &    3.28$\times$10$^{+00}$    &    8.79$\times$10$^{+01}$    &    1.57$\times$10$^{+01}$   \\
1.81$\times$10$^{+01}$    &    7.11$\times$10$^{+01}$    &    2.61$\times$10$^{+00}$    &    7.74$\times$10$^{+01}$    &    1.34$\times$10$^{+01}$    \\
1.85$\times$10$^{+01}$    &    6.46$\times$10$^{+01}$    &    2.18$\times$10$^{+00}$    &    7.01$\times$10$^{+01}$    &    1.19$\times$10$^{+01}$    \\
1.91$\times$10$^{+01}$    &    5.78$\times$10$^{+01}$   &    1.72$\times$10$^{+00}$    &    6.26$\times$10$^{+01}$    &    1.03$\times$10$^{+01}$    \\
1.95$\times$10$^{+01}$    &    5.28$\times$10$^{+01}$    &    1.41$\times$10$^{+00}$    &    5.70$\times$10$^{+01}$    &    8.95$\times$10$^{+00}$    \\
1.97$\times$10$^{+01}$   &    5.08$\times$10$^{+01}$    &    1.27$\times$10$^{+00}$    &    5.47$\times$10$^{+01}$    &    8.36$\times$10$^{+00}$    \\
2.00$\times$10$^{+01}$    &    4.65$\times$10$^{+01}$    &    1.12$\times$10$^{+00}$    &    4.98$\times$10$^{+01}$    &    6.98$\times$10$^{+00}$    \\
2.03$\times$10$^{+01}$    &    4.43$\times$10$^{+01}$    &    1.06$\times$10$^{+00}$    &    4.73$\times$10$^{+01}$   &    6.43$\times$10$^{+00}$    \\
2.05$\times$10$^{+01}$   &    4.34$\times$10$^{+01}$    &    1.04$\times$10$^{+00}$   &    4.63$\times$10$^{+01}$   &    6.19$\times$10$^{+00}$    \\
2.12$\times$10$^{+01}$    &    3.98$\times$10$^{+01}$    &    9.60$\times$10$^{-01}$    &    4.22$\times$10$^{+01}$    &    5.37$\times$10$^{+00}$    \\
2.15$\times$10$^{+01}$   &    3.83$\times$10$^{+01}$    &    9.30$\times$10$^{-01}$   &    4.06$\times$10$^{+01}$    &    5.08$\times$10$^{+00}$    \\
2.22$\times$10$^{+01}$    &    3.62$\times$10$^{+01}$   &    8.88$\times$10$^{-01}$    &    3.84$\times$10$^{+01}$    &    4.89$\times$10$^{+00}$    \\
2.24$\times$10$^{+01}$   &    3.66$\times$10$^{+01}$    &    9.09$\times$10$^{-01}$    &    3.89$\times$10$^{+01}$    &    5.23$\times$10$^{+00}$ \\
2.26$\times$10$^{+01}$   &    3.65$\times$10$^{+01}$    &    9.06$\times$10$^{-01}$   &    3.88$\times$10$^{+01}$    &    5.32$\times$10$^{+00}$    \\
2.27$\times$10$^{+01}$    &    3.66$\times$10$^{+01}$    &    9.08$\times$10$^{-01}$    &    3.89$\times$10$^{+01}$    &    5.42$\times$10$^{+00}$    \\
2.33$\times$10$^{+01}$    &    3.57$\times$10$^{+01}$    &    8.78$\times$10$^{-01}$    &    3.80$\times$10$^{+01}$    &    5.55$\times$10$^{+00}$    \\
2.38$\times$10$^{+01}$    &    3.47$\times$10$^{+01}$    &    8.63$\times$10$^{-01}$    &    3.70$\times$10$^{+01}$    &    5.40$\times$10$^{+00}$   \\
2.42$\times$10$^{+01}$    &    3.41$\times$10$^{+01}$   &    8.30$\times$10$^{-01}$    &    3.63$\times$10$^{+01}$    &    5.40$\times$10$^{+00}$    \\
2.51$\times$10$^{+01}$    &    3.26$\times$10$^{+01}$    &    7.88$\times$10$^{-01}$    &    3.47$\times$10$^{+01}$   &    5.14$\times$10$^{+00}$    \\
2.69$\times$10$^{+01}$    &    2.91$\times$10$^{+01}$    &    6.90$\times$10$^{-01}$    &    3.09$\times$10$^{+01}$    &    4.47$\times$10$^{+00}$    \\
2.76$\times$10$^{+01}$    &    2.82$\times$10$^{+01}$    &    6.60$\times$10$^{-01}$    &    2.99$\times$10$^{+01}$    &    4.31$\times$10$^{+00}$    \\
3.09$\times$10$^{+01}$    &    2.19$\times$10$^{+01}$    &    4.51$\times$10$^{-01}$    &    2.33$\times$10$^{+01}$   &    3.57$\times$10$^{+00}$    \\
3.45$\times$10$^{+01}$   &    1.67$\times$10$^{+01}$   &    3.27$\times$10$^{-01}$    &    1.80$\times$10$^{+01}$    &    3.29$\times$10$^{+00}$    \\

\hline
\hline
  \end{tabular}
 \label{tab:EvaluatedCSResults}
\end{table}

\section{\label{sec::level4} Discussion}

We first discuss the issues regarding simultaneous reproduction of the $^{96}{\rm Zr}(\alpha,n)$ cross section and $^{96}{\rm Zr}(\alpha,\alpha)$ differential cross section before turning to the evaluated cross section, associated astrophysical reaction rate, and implications for medical isotope production.

\subsection{Evaluated cross section results}

When examining the 95\% confidence interval contours of sub-figures (a) and (b) of Figure~\ref{fig:chi2_CS}, it is apparent that the corresponding contours are consistent but more restrictive for the $\alpha$-optical potential parameters when including the data of Ref.~\cite{Kiss21}. This is likely due to the small uncertainties of that data set, along with the larger $E_{\alpha}$ range that the calculations must reproduce. It is unsurprising that the contours are more similar for the nuclear level density parameters, as these impact the calculated cross section above the energy range covered by Ref.~\cite{Kiss21}.

Tension arises when comparing the 95\% confidence interval contours of Figure~\ref{fig:chi2_CS} to those of Figure~\ref{fig:chi2_diffCS}. When including the data of Ref.~\cite{Kiss21} (Figure~\ref{fig:chi2_CS}a), reproducing the $^{96}{\rm Zr}(\alpha,n)$ cross section data requires both $W$ and $a_{v}$ to be lower than calculations that successfully reproduce the $^{96}{\rm Zr}(\alpha,\alpha)$ differential cross section data. When excluding the data of Ref.~\cite{Kiss21} (Figure~\ref{fig:chi2_CS}b), the tension with $W$ is relieved and the 95\% confidence interval contours overlap for almost all of the panels. However, though the the optimal regions in the $r_{v}$ versus $a_{v}$ phase-space are closer than for the case when Ref.~\cite{Kiss21} data are included, they nonetheless still do not overlap. This indicates that the $\alpha$-optical potential of Ref.~\cite{mcfadden} is inadequate to fully reproduce measured data from the $^{96}{\rm Zr}$+$\alpha$ reaction. Figure~\ref{fig:diffCS} demonstrates that the ATOMKI-V2 potential, which has recently been employed for nuclear astrophysics studies~\cite{Kiss21,Szeg21,Psal22}, is similarly challenged. Calculations using the ATOMKI-V2 potential also do not reproduce the $^{96}{\rm Zr}(\alpha,n)$ cross section data for $E_{\alpha}\gtrsim$13~MeV; however, this potential has been optimized for sub-Coulomb barrier energies and therefore such a discrepancy is not unexpected~\cite{Mohr20}.

Inspired by Refs.~\cite{Nolt87,Avri94}, we performed exploratory calculations expanding our $\alpha$-optical potential parameter phase space to include a linear energy dependence for $V$ and $W$. This addition did not resolve the tension in $r_{v}$ versus $a_{v}$ and thus we do not discuss these exploratory calculations further. We did not investigate higher-order energy-dependencies (e.g. as in Ref.~\cite{Avri14}), nor independent variations of $r_{w}$ and $a_{w}$, as we desired to maintain a relatively simple functional form, given the limited data set involved in the model-experiment comparisons.

\begin{figure*}[htb]
\begin{center}
\includegraphics[height=12cm,width=2\columnwidth]{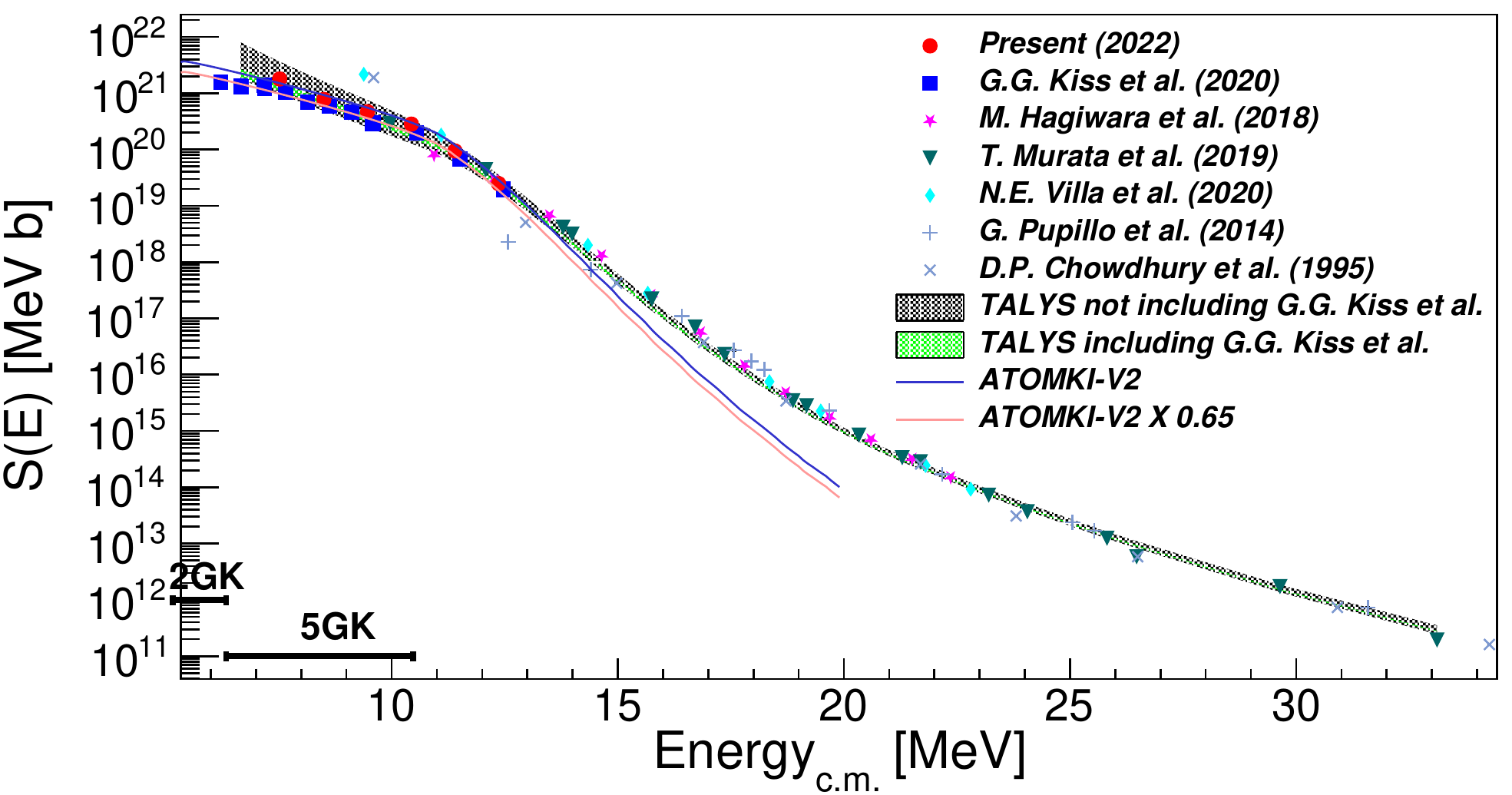}%
\begin{picture}(0,0)
\put(-420,120){\includegraphics[height=4.05cm]{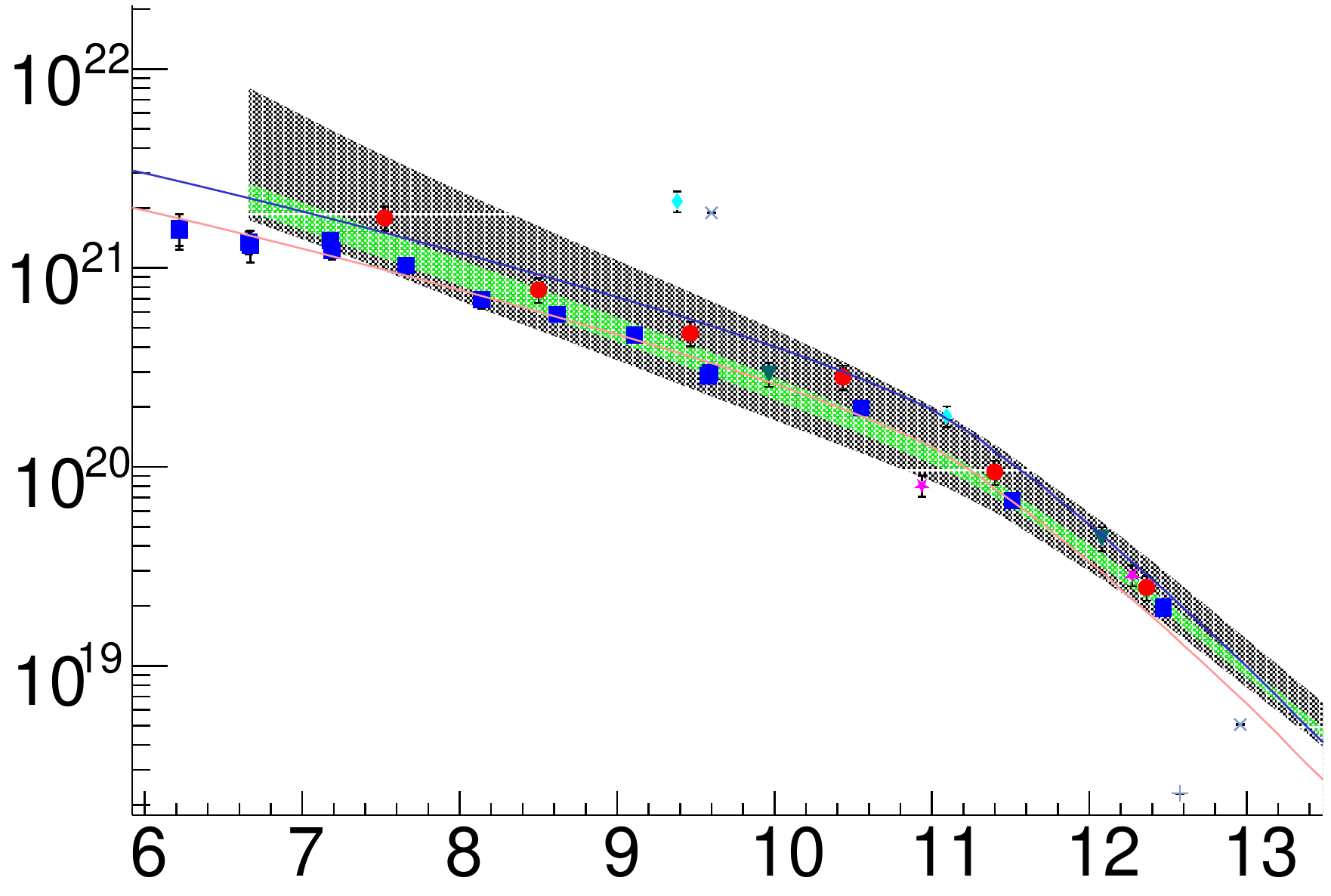}} 
\end{picture}
\caption{S-factor corresponding to the cross section data shown in Figure~\ref{fig:CSbands}. Bars in the lower-left indicate the astrophysical window, i.e. the region that contributes between 10-90\% of the integrand in the calculation of the astrophysical reaction rate, at the indicated temperature. The inset highlights the low-energy region.}
\label{fig:Sfactor}
\end{center}
\end{figure*}

Instead, we focus on calculation results that best reproduced the $^{96}{\rm Zr}(\alpha,n)$ cross section. These are the {\tt Talys} calculations for which all parameters are located within all of the 95\% confidence interval contours of Figure~\ref{fig:chi2_CS}a or \ref{fig:chi2_CS}b. The former are represented by the green bands of Figures~\ref{fig:CSbands}--\ref{fig:Activity}, while the latter are represented by the black bands within those figures. We also compare our results to the Hauser-Feshbach calculations performed using the ATOMKI-V2 $\alpha$-optical potential~\cite{Mohr20}, with the empirical scaling ($\times0.65$) adopted by Ref.~\cite{Kiss21}.

A more detailed comparison between our experimental results, Hauser-Feshbach calculation results, and prior results from the literature is enabled by considering the the S-factor $S(E)$. This removes the trivial energy dependencies of the cross section due to geometry (from the de Broglie wavelength) and the Coulomb barrier.
    $S(E)=\sigma(E)E\exp\left(2\pi\eta\right)$,
where $E$ is the center-of-mass energy and $\eta$ is the Sommerfeld parameter. The latter is
    $\eta=\alpha_{\rm fs}Z_{1}Z_{2}\sqrt{\frac{\mu c^{2}}{2E}}$,
where $\alpha_{\rm fs}$ is the fine-structure constant and $\mu=(m_{1}m_{2})/(m_{1}+m_{2})$ is the reduced mass of the reactants with masses $m_{i}$ and nuclear charges $Z_{i}$.

Figure~\ref{fig:Sfactor} shows $S(E)$ corresponding to the cross section data of Figure~\ref{fig:CSbands}. In general, uncertainty bands encompass the experimental data for $^{96}{\rm Zr}(\alpha,n)$. When excluding the data of Ref.~\cite{Kiss21}, our calculations result in a larger $S(E)$ at the lowest energies. When the data of Ref.~\cite{Kiss21} are included, our calculated $S(E)$ adequately describe these data, demonstrating that the simple potential of Ref.~\cite{mcfadden} is adequate to describe these cross section data.

One possible origin of the discrepancy between our data and the results of Ref.~\cite{Kiss21} is the different approaches our works take to account for $\gamma$-summing in the activation measurements. We employ the correction method of Ref.~\cite{SEMKOW1990437} that is briefly described in Section~\ref{sec::level2}. Instead, Ref.~\cite{Kiss21} performed relative measurements of activated samples in near and far geometries in order to empirically calibrate the summing correction for their near-geometry measurements that were performed for their lowest activation energies. We stress that both $\gamma$-summing correction techniques are relatively standard and there is no strong reason to favor one over the other. Another difference between the measurements is the use of an aluminum backing for the very thin targets of Ref.~\cite{Kiss21}, as opposed to the free-standing targets used here. In principle, long-lived species created by reactions of the $\alpha$ beam on contaminants in the backing could complicate the $\gamma$-background subtraction. Given the importance of the low-energy region for nuclear astrophysics, described in the following subsection, the discrepancy presented here is a strong motivation for independent follow-up measurements for $E_{\alpha}\leq8$~MeV.

\subsection{Implications for astrophysics}

\begin{figure}[]
\begin{center}
\includegraphics[width=\columnwidth]{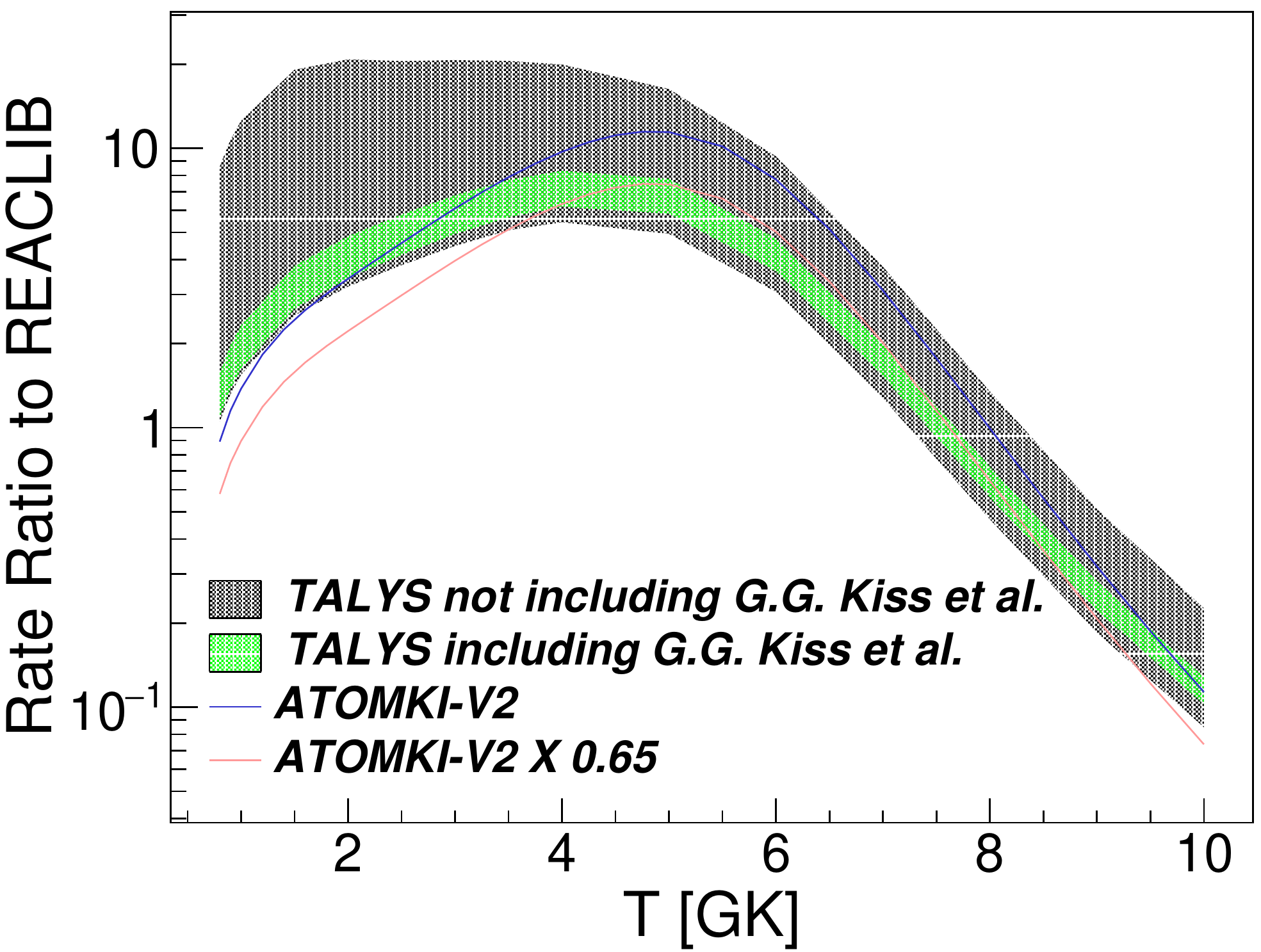}
\caption{Ratio of the $^{96}{\rm Zr}(\alpha,n)$ astrophysical reaction rates calculated in this work and reported in Ref.~\cite{Kiss21} to the rate of Ref.~\cite{Raus00}, which is recommended in the ReacLib database~\cite{Cybu10}.}
\label{fig:ReactionRate}
\end{center}
\end{figure}

As described in Section~\ref{sec:level1}, the $^{96}{\rm Zr}(\alpha,n)$ reaction rate at temperatures of $\approx$2--5~GK plays an important role in neutron-rich $\nu$-driven wind nucleosynthesis in CCSN. The corresponding energy region of interest for $S(E)$ is determined by considering the integrand of the astrophysical reaction rate,
$\langle\sigma v\rangle \propto \int_{0}^{\infty}S(E)\exp\left(-E/(k_{B}T) -2\pi\eta\right)dE$,
where $k_{B}$ is the Boltzmann constant and $T$ is the astrophysical environment temperature. We refer to the energy-region contributing between 10--90\% of the area of the integrand as the astrophysical window. Figure~\ref{fig:Sfactor} shows the astrophysical window for temperatures of interest for this work. It is apparent that the astrophysical reaction rate depends on $S(E)$ in the energy region that is primarily constrained by our experimental results and those of Ref.~\cite{Kiss21}. The rate below 6~GK, shown in Figure~\ref{fig:ReactionRate}, is particularly sensitive to the discrepancy between our work and Ref.~\cite{Kiss21}. 

\begin{table}[h!]
\centering
\caption{Astrophysical reaction rates N$_{A}$$\langle$$\sigma$v$\rangle$ for $^{96}$Zr($\alpha$,n)$^{99}$Mo in units of cm$^{3}$ s$^{-1}$ mole$^{-1}$ and temperature in GK (T9). The best-fit and upper(lower) rates are reported for the evaluated results that include (rate A) or exclude (rate B) Ref.~\cite{Kiss21}, along with uncertainties (unc.).\label{tab:RateResults}}
\begin{tabular}{@{}ccccc@{}}
\hline
\hline
T9  & Rate A & unc. A & Rate B & unc. B \\
\hline
0.8  &    1.46$\times$10$^{-31}$    &    3.96$\times$10$^{-32}$   &    8.55$\times$10$^{-31}$    &    7.50$\times$10$^{-31}$    \\
0.9  &    6.06$\times$10$^{-28}$    &    1.65$\times$10$^{-28}$    &    3.53$\times$10$^{-27}$    &    3.09$\times$10$^{-27}$   \\
1.0  &    4.74$\times$10$^{-25}$    &    1.29$\times$10$^{-25}$   &    2.75$\times$10$^{-24}$    &    2.41$\times$10$^{-24}$    \\
1.1  &    5.78$\times$10$^{-17}$   &    1.58$\times$10$^{-17}$    &    3.13$\times$10$^{-16}$    &    2.72$\times$10$^{-16}$   \\
1.2  &    1.16$\times$10$^{-16}$   &    3.15$\times$10$^{-17}$    &    6.27$\times$10$^{-16}$    &    5.43$\times$10$^{-16}$    \\
1.3  &    1.73$\times$10$^{-16}$   &    4.73$\times$10$^{-17}$   &    9.40$\times$10$^{-16}$    &    8.15$\times$10$^{-16}$    \\
1.4  &    2.31$\times$10$^{-16}$    &    6.30$\times$10$^{-17}$    &    1.25$\times$10$^{-15}$    &    1.09$\times$10$^{-15}$   \\
1.5  &    2.89$\times$10$^{-16}$    &    7.88$\times$10$^{-17}$    &    1.57$\times$10$^{-15}$    &    1.36$\times$10$^{-15}$    \\
1.6  &    2.09$\times$10$^{-12}$    &    5.58$\times$10$^{-13}$    &    9.72$\times$10$^{-12}$    &    8.19$\times$10$^{-12}$    \\
1.7  &    4.18$\times$10$^{-12}$    &    1.12$\times$10$^{-12}$    &    1.94$\times$10$^{-11}$    &    1.64$\times$10$^{-11}$    \\
1.8  &    6.28$\times$10$^{-12}$    &    1.67$\times$10$^{-12}$    &    2.92$\times$10$^{-11}$    &    2.46$\times$10$^{-11}$    \\
1.9  &    8.37$\times$10$^{-12}$   &    2.23$\times$10$^{-12}$    &    3.89$\times$10$^{-11}$    &    3.27$\times$10$^{-11}$    \\
2.0  &    1.05$\times$10$^{-11}$    &    2.79$\times$10$^{-12}$    &    4.86$\times$10$^{-11}$   &    4.09$\times$10$^{-11}$    \\
2.1  &    1.69$\times$10$^{-09}$    &    4.33$\times$10$^{-10}$    &    6.24$\times$10$^{-09}$    &    4.99$\times$10$^{-09}$    \\
2.2  &    3.37$\times$10$^{-09}$    &    8.63$\times$10$^{-10}$    &    1.24$\times$10$^{-08}$   &    9.94$\times$10$^{-09}$    \\
2.3  &    5.04$\times$10$^{-09}$    &    1.29$\times$10$^{-09}$    &    1.86$\times$10$^{-08}$    &    1.49$\times$10$^{-08}$   \\
2.4  &    6.72$\times$10$^{-09}$    &    1.72$\times$10$^{-09}$   &    2.48$\times$10$^{-08}$    &    1.98$\times$10$^{-08}$    \\
2.5  &    8.40$\times$10$^{-09}$    &    2.15$\times$10$^{-09}$    &    3.10$\times$10$^{-08}$    &    2.48$\times$10$^{-08}$    \\
2.6  &    2.13$\times$10$^{-07}$    &    5.22$\times$10$^{-08}$   &    6.08$\times$10$^{-07}$    &    4.57$\times$10$^{-07}$    \\
2.7  &    4.17$\times$10$^{-07}$    &    1.02$\times$10$^{-07}$   &    1.18$\times$10$^{-06}$    &    8.89$\times$10$^{-07}$    \\
2.8  &    6.22$\times$10$^{-07}$    &    1.52$\times$10$^{-07}$   &    1.76$\times$10$^{-06}$    &    1.32$\times$10$^{-06}$    \\
2.9  &    8.26$\times$10$^{-07}$   &    2.02$\times$10$^{-07}$   &    2.34$\times$10$^{-06}$   &    1.75$\times$10$^{-06}$   \\
3.0  &    1.03$\times$10$^{-06}$   &    2.52$\times$10$^{-07}$    &    2.92$\times$10$^{-06}$   &    2.19$\times$10$^{-06}$    \\
3.1  &    9.29$\times$10$^{-06}$   &    2.18$\times$10$^{-06}$    &    2.09$\times$10$^{-05}$    &    1.47$\times$10$^{-05}$    \\
3.2  &    1.75$\times$10$^{-05}$   &    4.10$\times$10$^{-06}$    &    3.89$\times$10$^{-05}$    &    2.71$\times$10$^{-05}$    \\
3.3  &    2.58$\times$10$^{-05}$    &    6.02$\times$10$^{-06}$   &    5.69$\times$10$^{-05}$   &    3.96$\times$10$^{-05}$   \\
3.4  &    3.41$\times$10$^{-05}$   &    7.95$\times$10$^{-06}$    &    7.49$\times$10$^{-05}$   &    5.21$\times$10$^{-05}$    \\
3.5  &    4.23$\times$10$^{-05}$    &    9.87$\times$10$^{-06}$   &    9.29$\times$10$^{-05}$    &    6.46$\times$10$^{-05}$    \\
3.6  &    2.02$\times$10$^{-04}$    &    4.55$\times$10$^{-05}$    &    3.74$\times$10$^{-04}$    &    2.43$\times$10$^{-04}$    \\
3.7  &    3.62$\times$10$^{-04}$    &    8.11$\times$10$^{-05}$    &    6.56$\times$10$^{-04}$    &    4.22$\times$10$^{-04}$    \\
3.8  &    5.22$\times$10$^{-04}$    &    1.17$\times$10$^{-04}$    &    9.37$\times$10$^{-04}$    &    6.01$\times$10$^{-04}$    \\
3.9  &    6.82$\times$10$^{-04}$    &    1.52$\times$10$^{-04}$   &    1.22$\times$10$^{-03}$    &    7.80$\times$10$^{-04}$    \\
4.0  &    8.42$\times$10$^{-04}$    &    1.88$\times$10$^{-04}$   &    1.50$\times$10$^{-03}$   &    9.59$\times$10$^{-04}$    \\
4.1  &    8.29$\times$10$^{-03}$    &    1.73$\times$10$^{-03}$    &    1.23$\times$10$^{-02}$    &    7.08$\times$10$^{-03}$    \\
4.2  &    1.57$\times$10$^{-02}$    &    3.26$\times$10$^{-03}$    &    2.30$\times$10$^{-02}$   &    1.32$\times$10$^{-02}$    \\
4.3  &    2.32$\times$10$^{-02}$    &    4.80$\times$10$^{-03}$    &    3.38$\times$10$^{-02}$    &    1.93$\times$10$^{-02}$    \\
4.4  &    3.06$\times$10$^{-02}$    &    6.34$\times$10$^{-03}$    &    4.46$\times$10$^{-02}$    &    2.54$\times$10$^{-02}$    \\
4.5  &    3.81$\times$10$^{-02}$    &    7.88$\times$10$^{-03}$    &    5.53$\times$10$^{-02}$    &    3.16$\times$10$^{-02}$    \\
4.6  &    4.55$\times$10$^{-02}$    &    9.41$\times$10$^{-03}$   &    6.61$\times$10$^{-02}$    &    3.77$\times$10$^{-02}$    \\
4.7  &    5.30$\times$10$^{-02}$    &    1.10$\times$10$^{-02}$   &    7.69$\times$10$^{-02}$    &    4.38$\times$10$^{-02}$    \\
4.8  &    6.04$\times$10$^{-02}$    &    1.25$\times$10$^{-02}$   &    8.76$\times$10$^{-02}$    &    4.99$\times$10$^{-02}$    \\
4.9  &    6.79$\times$10$^{-02}$    &    1.40$\times$10$^{-02}$    &    9.84$\times$10$^{-02}$    &    5.60$\times$10$^{-02}$    \\
5.0  &    7.53$\times$10$^{-02}$    &    1.56$\times$10$^{-02}$    &    1.09$\times$10$^{-01}$    &    6.21$\times$10$^{-02}$    \\
6.0  &    1.50$\times$10$^{+00}$    &    2.94$\times$10$^{-01}$    &    2.05$\times$10$^{+00}$    &    1.11$\times$10$^{+00}$    \\
7.0  &    9.85$\times$10$^{+00}$    &    1.89$\times$10$^{+00}$    &    1.32$\times$10$^{+01}$    &    6.88$\times$10$^{+00}$    \\
8.0  &    3.33$\times$10$^{+01}$    &    6.31$\times$10$^{+00}$    &    4.39$\times$10$^{+01}$    &    2.22$\times$10$^{+01}$    \\
9.0  &    7.73$\times$10$^{+01}$    &    1.46$\times$10$^{+01}$    &    1.00$\times$10$^{+02}$    &    4.90$\times$10$^{+01}$    \\
10.0   &   1.43$\times$10$^{+02}$    &    2.67$\times$10$^{+01}$    &    1.83$\times$10$^{+02}$    &    8.51$\times$10$^{+01}$    \\

\hline
\hline
  \end{tabular}
 \label{table:RateResults}
\end{table}

The astrophysical reaction rates calculated with {\tt Talys} from our best-fits, along with the reaction rate uncertainty bands that encompass results within the 95\% confidence interval contours of Figure~\ref{fig:chi2_CS}, are reported in Table~\ref{tab:RateResults}.

When we include the experimental data of Ref.~\cite{Kiss21} in our evaluation, our reaction rate is nearly in agreement with the results from that work (reaching $\approx\times$2 disagreement at 2~GK), with an uncertainty between $\approx30-200$\% in the temperature range of interest. If we exclude the experimental data of Ref.~\cite{Kiss21} in our reaction rate evaluation, we find that the $^{96}{\rm Zr}(\alpha,n)$ reaction rate could be up to 10$\times$ larger than the rate reported in Ref.~\cite{Kiss21} and up to 20$\times$ higher than adopted in ReacLib~\cite{Cybu10} within the temperature range of interest. Based on the reaction rate sensitivity study results of Ref.~\cite{bliss2020} (see their Figure~3), this could result in a roughly 0.4~dex increase in the predicted abundance of silver isotopes~\footnote{As Ref.~\cite{bliss2020} shows, the impact of an individual reaction rate depends on the rates adopted for many nuclear reactions. For our estimated impact, we are concentrating on the average linear trend of their silver abundance versus $^{96}{\rm Zr}(\alpha,n)$ rate variation scatter plot.}. This is to be compared to an achievable observational uncertainty of around 0.2~dex for silver abundances in metal poor stars~\cite[e.g.][]{Roed12}, where the majority of silver in these objects is thought to come from weak $r$-process nucleosynthesis~\cite{Hans12}.

\subsection{Implications for medical physics}

The two primary concerns in accelerator-based medical isotope production are the yield of the species of interest and the yield of radioactive contaminants, where the latter can contribute unnecessary dose to radiation workers and patients. Contaminants of the same element as the isotope of interest are of particular concern, as these are not removed by chemical separation methods~\cite{Lame19}. The yield of a specific isotope from a nuclear reaction is calculated by $Y=\int_{E_{\rm exit}}^{E_{\rm entr}}(\sigma(E)/\mathbb{S}(E))dE$, where $\mathbb{S}(E)$ is the stopping power of the beam in the target and $E_{\rm entr}$ and $E_{\rm exit}$ are the energies at which the beam enters and exits the target, with $E_{\rm exit}=0$ for stopping targets.

\begin{figure}[t]
\begin{center}
\includegraphics[height=6.5cm,width=\columnwidth]{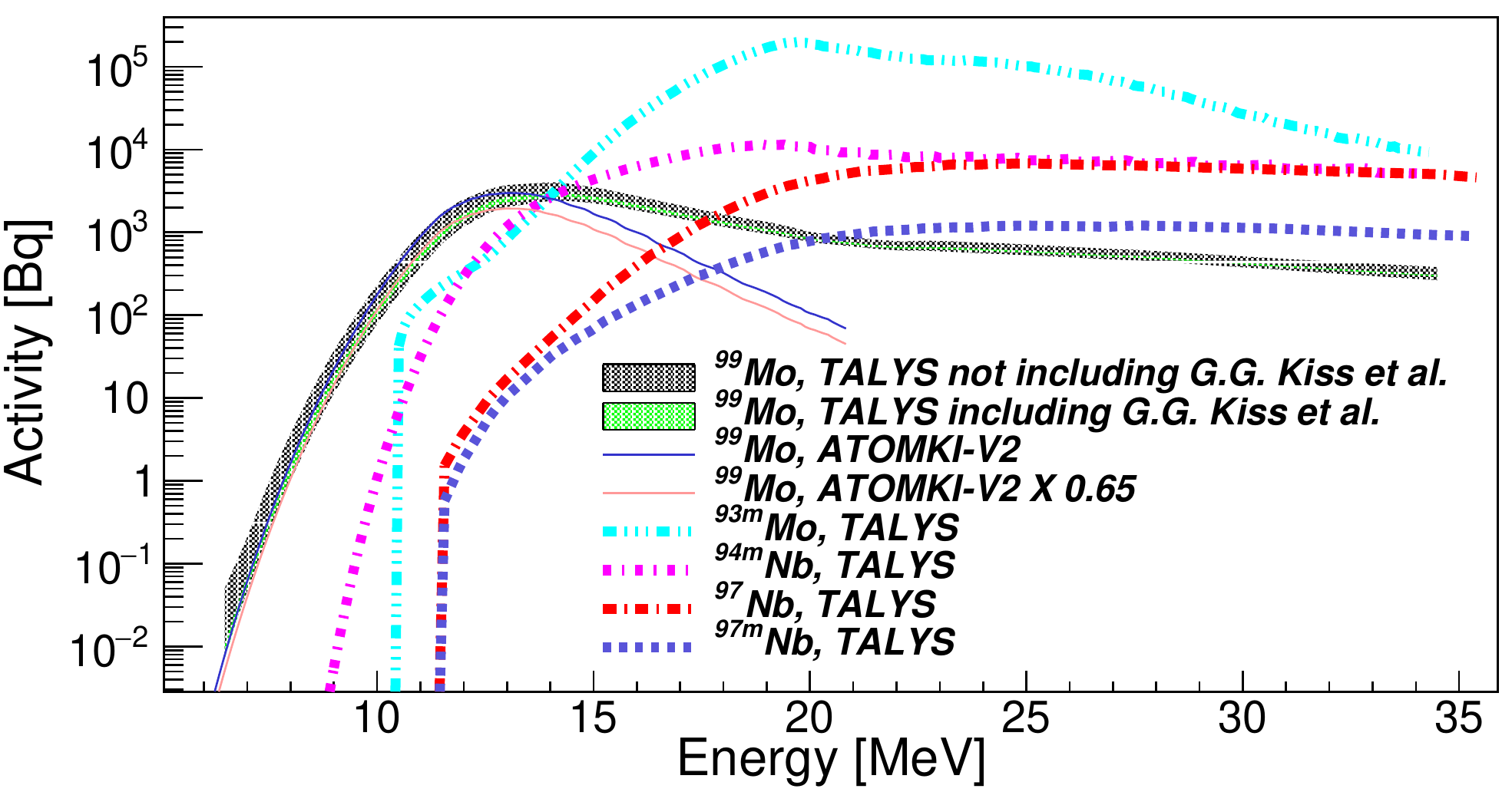}
\caption{Activity produced after 3~hr of irradiation of a 1~mg/cm$^{2}$ natural zirconium target with 1~$\mu$A of $\alpha$-particles. The $^{99}{\rm Mo}$ activity is calculated using the Hauser-Feshbach calculation results of Figure~\ref{fig:CSbands}. The activity for all other species relies on cross sections calculated with the {\tt Talys} default parameters.}
\label{fig:Activity}
\end{center}
\end{figure}

To assess the impact of our results, we performed activation calculations for helium ions impinging on a natural zirconium target, adopting the Hauser-Feshbach calculation results of Figure~\ref{fig:CSbands} for the $^{96}{\rm Zr}(\alpha,n)$ cross section and results from {\tt Talys} calculations performed with default settings otherwise. For $\mathbb{S}(E)$, we use the calculations of {\tt SRIM 2013}~\cite{Zieg10}, which are in agreement with the only published stopping powers of helium in zirconium~\cite{Lin73,Mont17} and have been found to reproduce measured stopping powers in elemental solids at these energies within 4\%~\cite{Paul05}.
Figure~\ref{fig:Activity} shows the results of these calculations, where we have adopted the somewhat arbitrary conditions of a 1~$\mu$A $\alpha$-beam impinging on a 1~mg/cm$^{2}$ natural zirconium target for 3~hr of irradiation.  We only report results for $E_{\alpha}\leq20$~MeV for species produced with an activity greater than 100~Bq within this energy window.

Aside from $^{99}{\rm Mo}$, the significant contaminants that are also produced for these irradiation conditions include $^{\rm 93m}{\rm Mo}$ ($t_{1/2}=6.9$~hr~\cite{Bagl11}), $^{\rm 94m}{\rm Nb}$ ($t_{1/2}=6.3$~m~\cite{Abri06}), $^{97}{\rm Nb}$ ($t_{1/2}=1.2$~hr~\cite{Nica10}), and $^{\rm 97m}{\rm Nb}$ ($t_{1/2}=59$~s~\cite{Nica10}). Though the $^{96}{\rm Zr}(\alpha,n)$ cross section has a maximum around $E_{\alpha}=15$~MeV, the activity contribution from contaminants is dominant above $E_{\alpha}=14$~MeV. These conclusions are essentially not changed for the various $^{96}{\rm Zr}(\alpha,n)$ cross sections included in Figure~\ref{fig:Activity}. The optimum $\alpha$-beam energy for $^{99}{\rm Mo}$ production is therefore somewhere below 14~MeV for a natural zirconium target, depending on contaminant tolerances. For instance, the $^{99}{\rm Mo}$/contaminant activity ratio first reaches $\times2$ around 12.7~MeV and $\times10$ around 10.5~MeV. For any $E_{\alpha}$, the yield of $^{99}{\rm Mo}$ is somewhat higher when adopting our results, in particular the calculations satisfying the 95\% confidence interval contours of Figure~\ref{fig:chi2_CS}b, as opposed to the Hauser-Feshbach results of Ref.~\cite{Kiss21}.

We note that Ref.~\cite{Vill20} recommends $E_{\alpha}=19.5$~MeV for $^{99}{\rm Mo}$ production via $^{96}{\rm Zr}(\alpha,n)$, but they assume the use of a highly-enriched  $^{96}{\rm Zr}$ target. For these conditions and a 44.1~mg/cm$^{2}$ target, they find a $^{99}{\rm Mo}$ yield of 1.5~MBq/$\mu$A for 1~hr of irradiation. For our evaluated cross sections and the same conditions, we find 0.96$\pm$0.052~MBq/$\mu$A and 1.07$\pm$0.25~MBq/$\mu$A when including or excluding the results of Ref.~\cite{Kiss21} in our cross section evaluation, respectively. The difference is due to our evaluated cross section falling slightly under the world data for $E_\alpha\approx15-20$~MeV.

\section{\label{sec:conclusion} Conclusions}
We performed measurements of the $^{96}{\rm Zr}(\alpha,n)$ cross section via the activation technique at the Edwards Accelerator Laboratory and performed large-scale Hauser-Feshbach calculations to evaluate the world data for this reaction. Our activation measurements are largely in agreement with previous high-precision results~\cite{Kiss21}, but we find a higher cross section, especially for our lowest-energy measurement point. Our large-scale Hauser-Feshbach calculations, which employ a simple $\alpha$-optical potential of the style in Ref.~\cite{mcfadden}, are not able to simultaneously consistently describe the world data for the $^{96}{\rm Zr}(\alpha,n)$ cross section and the $^{96}{\rm Zr}(\alpha,\alpha)$ differential cross section, similar to the ATOMKI-V2 potential. However, we find that the simple $\alpha$-optical potential is adequate to describe the $^{96}{\rm Zr}(\alpha,n)$ cross section data. This indicates that phenomenological optical potentials assuming a Woods-Saxon form may yet be adequate to describe $(\alpha,n)$ reactions at low energies in this mass region, albeit with regionally-adjusted model parameters, contrary to the suggestions of Ref.~\cite{Kiss21,Szeg21}. Further investigations will be needed in this region of the nuclear landscape to identify the best adjusted parameterization. Special attention should be paid to the tail of the $\alpha$-optical potential, as this is what determines the $(\alpha,n)$ cross section at low energies.

We present newly evaluated $^{96}{\rm Zr}(\alpha,n)$ cross sections, along with corresponding astrophysical reaction rates at temperatures relevant for $\nu$-driven wind nucleosynthesis in CCSN. The larger low-energy cross section found in this work relative to Ref.~\cite{Kiss21} allows for a correspondingly larger astrophysical reaction rate. This would likely enhance the production of silver in neutron-rich $\nu$-driven winds. Given this discrepancy, we encourage additional high-precision measurements of the $^{96}{\rm Zr}(\alpha,n)$ cross section at $E_{\alpha}\leq8$~MeV.

We also present results from calculations of $^{99}{\rm Mo}$ production for hypothetical medical isotope production scenarios. We find that the optimum irradiation energy is $E_{\alpha}<14$~MeV for a natural zirconium target, where the optimum energy depends on tolerances for co-producing radioactive contaminants. For an isotopically enriched $^{96}{\rm Zr}$ target, the $^{99}{\rm Mo}$ activity resulting from a hypothetical irradiation scenario is around 30\% smaller than recent results from Ref.~\cite{Vill20}. However, our evaluated cross section results generally lay below the world data for $E_{\alpha}\approx15-20$~MeV, where the majority of the yield comes from in this hypothetical scenario, given the challenges of reproducing this energy range with Hauser-Feshbach calculations. It is possible that measurements of $\rho$ for $^{99}{\rm Mo}$ would improve the Hauser-Feshbach description of this energy region, though a more sophisticated $\alpha$-optical potential may be required.

\begin{acknowledgements}
 We thank Peter Mohr for assistance in performing Hauser-Feshbach calculations using the ATOMKI-V2 potential. This work was supported in part by the U.S. Department of Energy Office of Science under Grants No. DE-FG02-88ER40387 and DE-SC0019042 and the U.S. National Nuclear Security Administration through Grants No. , DE-NA0003883 and DE-NA0003909. The helium ion source was provided by Grant No. PHY-1827893 from the U.S. National Science Foundation. We also benefited from support by the U.S. National Science Foundation under Grant No. PHY-1430152 (Joint Institute for Nuclear Astrophysics -- Center for the Evolution of the Elements). The Zr target used in this research was supplied by the U.S. Department of Energy Office of Science by the Isotope Program in the Office of Nuclear Physics.
\end{acknowledgements}
\bibliographystyle{apsrev4-2}
\bibliography{references}

\begin{thebibliography}{57}%
\makeatletter
\providecommand \@ifxundefined [1]{%
 \@ifx{#1\undefined}
}%
\providecommand \@ifnum [1]{%
 \ifnum #1\expandafter \@firstoftwo
 \else \expandafter \@secondoftwo
 \fi
}%
\providecommand \@ifx [1]{%
 \ifx #1\expandafter \@firstoftwo
 \else \expandafter \@secondoftwo
 \fi
}%
\providecommand \natexlab [1]{#1}%
\providecommand \enquote  [1]{``#1''}%
\providecommand \bibnamefont  [1]{#1}%
\providecommand \bibfnamefont [1]{#1}%
\providecommand \citenamefont [1]{#1}%
\providecommand \href@noop [0]{\@secondoftwo}%
\providecommand \href [0]{\begingroup \@sanitize@url \@href}%
\providecommand \@href[1]{\@@startlink{#1}\@@href}%
\providecommand \@@href[1]{\endgroup#1\@@endlink}%
\providecommand \@sanitize@url [0]{\catcode `\\12\catcode `\$12\catcode
  `\&12\catcode `\#12\catcode `\^12\catcode `\_12\catcode `\%12\relax}%
\providecommand \@@startlink[1]{}%
\providecommand \@@endlink[0]{}%
\providecommand \url  [0]{\begingroup\@sanitize@url \@url }%
\providecommand \@url [1]{\endgroup\@href {#1}{\urlprefix }}%
\providecommand \urlprefix  [0]{URL }%
\providecommand \Eprint [0]{\href }%
\providecommand \doibase [0]{https://doi.org/}%
\providecommand \selectlanguage [0]{\@gobble}%
\providecommand \bibinfo  [0]{\@secondoftwo}%
\providecommand \bibfield  [0]{\@secondoftwo}%
\providecommand \translation [1]{[#1]}%
\providecommand \BibitemOpen [0]{}%
\providecommand \bibitemStop [0]{}%
\providecommand \bibitemNoStop [0]{.\EOS\space}%
\providecommand \EOS [0]{\spacefactor3000\relax}%
\providecommand \BibitemShut  [1]{\csname bibitem#1\endcsname}%
\let\auto@bib@innerbib\@empty
\bibitem [{\citenamefont {Bliss}\ \emph {et~al.}(2020)\citenamefont {Bliss},
  \citenamefont {Arcones}, \citenamefont {Montes},\ and\ \citenamefont
  {Pereira}}]{bliss2020}%
  \BibitemOpen
  \bibfield  {author} {\bibinfo {author} {\bibfnamefont {J.}~\bibnamefont
  {Bliss}}, \bibinfo {author} {\bibfnamefont {A.}~\bibnamefont {Arcones}},
  \bibinfo {author} {\bibfnamefont {F.}~\bibnamefont {Montes}},\ and\ \bibinfo
  {author} {\bibfnamefont {J.}~\bibnamefont {Pereira}},\ }\href
  {https://doi.org/10.1103/PhysRevC.101.055807} {\bibfield  {journal} {\bibinfo
   {journal} {Phys. Rev. C}\ }\textbf {\bibinfo {volume} {101}},\ \bibinfo
  {pages} {055807} (\bibinfo {year} {2020})}\BibitemShut {NoStop}%
\bibitem [{\citenamefont {Hagiwara}\ \emph {et~al.}(2018)\citenamefont
  {Hagiwara}, \citenamefont {Yashima}, \citenamefont {Sanami},\ and\
  \citenamefont {Yonai}}]{Hagi18}%
  \BibitemOpen
  \bibfield  {author} {\bibinfo {author} {\bibfnamefont {M.}~\bibnamefont
  {Hagiwara}}, \bibinfo {author} {\bibfnamefont {H.}~\bibnamefont {Yashima}},
  \bibinfo {author} {\bibfnamefont {T.}~\bibnamefont {Sanami}},\ and\ \bibinfo
  {author} {\bibfnamefont {S.}~\bibnamefont {Yonai}},\ }\href
  {https://doi.org/10.1007/s10967-018-6118-4} {\bibfield  {journal} {\bibinfo
  {journal} {J. Rad. Nuc. Chem.}\ }\textbf {\bibinfo {volume} {318}},\ \bibinfo
  {pages} {569} (\bibinfo {year} {2018})}\BibitemShut {NoStop}%
\bibitem [{\citenamefont {{Montes}}\ \emph {et~al.}(2007)\citenamefont
  {{Montes}}, \citenamefont {Beers}, \citenamefont {Cowan}, \citenamefont
  {Elliot}, \citenamefont {Farouqi}, \citenamefont {Gallino}, \citenamefont
  {Heil}, \citenamefont {Kratz}, \citenamefont {Pfeiffer}, \citenamefont
  {Pigatari},\ and\ \citenamefont {Schatz}}]{Mont07}%
  \BibitemOpen
  \bibfield  {author} {\bibinfo {author} {\bibfnamefont {F.}~\bibnamefont
  {{Montes}}}, \bibinfo {author} {\bibfnamefont {T.~C.}\ \bibnamefont {Beers}},
  \bibinfo {author} {\bibfnamefont {J.}~\bibnamefont {Cowan}}, \bibinfo
  {author} {\bibfnamefont {T.}~\bibnamefont {Elliot}}, \bibinfo {author}
  {\bibfnamefont {K.}~\bibnamefont {Farouqi}}, \bibinfo {author} {\bibfnamefont
  {R.}~\bibnamefont {Gallino}}, \bibinfo {author} {\bibfnamefont
  {M.}~\bibnamefont {Heil}}, \bibinfo {author} {\bibfnamefont {K.-L.}\
  \bibnamefont {Kratz}}, \bibinfo {author} {\bibfnamefont {B.}~\bibnamefont
  {Pfeiffer}}, \bibinfo {author} {\bibfnamefont {M.}~\bibnamefont {Pigatari}},\
  and\ \bibinfo {author} {\bibfnamefont {H.}~\bibnamefont {Schatz}},\ }\href
  {https://doi.org/10.1086/523084} {\bibfield  {journal} {\bibinfo  {journal}
  {\apj}\ }\textbf {\bibinfo {volume} {671}},\ \bibinfo {pages} {1685}
  (\bibinfo {year} {2007})}\BibitemShut {NoStop}%
\bibitem [{\citenamefont {{Sakari}}\ \emph {et~al.}(2018)\citenamefont
  {{Sakari}}, \citenamefont {{Placco}}, \citenamefont {{Farrell}},
  \citenamefont {{Roederer}}, \citenamefont {{Wallerstein}}, \citenamefont
  {{Beers}}, \citenamefont {{Ezzeddine}}, \citenamefont {{Frebel}},
  \citenamefont {{Hansen}}, \citenamefont {{Holmbeck}}, \citenamefont
  {{Sneden}}, \citenamefont {{Cowan}}, \citenamefont {{Venn}}, \citenamefont
  {{Davis}}, \citenamefont {{Matijevi{\v{c}}}}, \citenamefont {{Wyse}},
  \citenamefont {{Bland-Hawthorn}}, \citenamefont {{Chiappini}}, \citenamefont
  {{Freeman}}, \citenamefont {{Gibson}}, \citenamefont {{Grebel}},
  \citenamefont {{Helmi}}, \citenamefont {{Kordopatis}}, \citenamefont
  {{Kunder}}, \citenamefont {{Navarro}}, \citenamefont {{Reid}}, \citenamefont
  {{Seabroke}}, \citenamefont {{Steinmetz}},\ and\ \citenamefont
  {{Watson}}}]{Saka18}%
  \BibitemOpen
  \bibfield  {author} {\bibinfo {author} {\bibfnamefont {C.~M.}\ \bibnamefont
  {{Sakari}}}, \bibinfo {author} {\bibfnamefont {V.~M.}\ \bibnamefont
  {{Placco}}}, \bibinfo {author} {\bibfnamefont {E.~M.}\ \bibnamefont
  {{Farrell}}}, \bibinfo {author} {\bibfnamefont {I.~U.}\ \bibnamefont
  {{Roederer}}}, \bibinfo {author} {\bibfnamefont {G.}~\bibnamefont
  {{Wallerstein}}}, \bibinfo {author} {\bibfnamefont {T.~C.}\ \bibnamefont
  {{Beers}}}, \bibinfo {author} {\bibfnamefont {R.}~\bibnamefont
  {{Ezzeddine}}}, \bibinfo {author} {\bibfnamefont {A.}~\bibnamefont
  {{Frebel}}}, \bibinfo {author} {\bibfnamefont {T.}~\bibnamefont {{Hansen}}},
  \bibinfo {author} {\bibfnamefont {E.~M.}\ \bibnamefont {{Holmbeck}}},
  \bibinfo {author} {\bibfnamefont {C.}~\bibnamefont {{Sneden}}}, \bibinfo
  {author} {\bibfnamefont {J.~J.}\ \bibnamefont {{Cowan}}}, \bibinfo {author}
  {\bibfnamefont {K.~A.}\ \bibnamefont {{Venn}}}, \bibinfo {author}
  {\bibfnamefont {C.~E.}\ \bibnamefont {{Davis}}}, \bibinfo {author}
  {\bibfnamefont {G.}~\bibnamefont {{Matijevi{\v{c}}}}}, \bibinfo {author}
  {\bibfnamefont {R.~F.~G.}\ \bibnamefont {{Wyse}}}, \bibinfo {author}
  {\bibfnamefont {J.}~\bibnamefont {{Bland-Hawthorn}}}, \bibinfo {author}
  {\bibfnamefont {C.}~\bibnamefont {{Chiappini}}}, \bibinfo {author}
  {\bibfnamefont {K.~C.}\ \bibnamefont {{Freeman}}}, \bibinfo {author}
  {\bibfnamefont {B.~K.}\ \bibnamefont {{Gibson}}}, \bibinfo {author}
  {\bibfnamefont {E.~K.}\ \bibnamefont {{Grebel}}}, \bibinfo {author}
  {\bibfnamefont {A.}~\bibnamefont {{Helmi}}}, \bibinfo {author} {\bibfnamefont
  {G.}~\bibnamefont {{Kordopatis}}}, \bibinfo {author} {\bibfnamefont
  {A.}~\bibnamefont {{Kunder}}}, \bibinfo {author} {\bibfnamefont
  {J.}~\bibnamefont {{Navarro}}}, \bibinfo {author} {\bibfnamefont
  {W.}~\bibnamefont {{Reid}}}, \bibinfo {author} {\bibfnamefont
  {G.}~\bibnamefont {{Seabroke}}}, \bibinfo {author} {\bibfnamefont
  {M.}~\bibnamefont {{Steinmetz}}},\ and\ \bibinfo {author} {\bibfnamefont
  {F.}~\bibnamefont {{Watson}}},\ }\href
  {https://doi.org/10.3847/1538-4357/aae9df} {\bibfield  {journal} {\bibinfo
  {journal} {\apj}\ }\textbf {\bibinfo {volume} {868}},\ \bibinfo {eid} {110}
  (\bibinfo {year} {2018})}\BibitemShut {NoStop}%
\bibitem [{\citenamefont {{Bethe}}\ and\ \citenamefont
  {{Wilson}}(1985)}]{Beth85}%
  \BibitemOpen
  \bibfield  {author} {\bibinfo {author} {\bibfnamefont {H.~A.}\ \bibnamefont
  {{Bethe}}}\ and\ \bibinfo {author} {\bibfnamefont {J.~R.}\ \bibnamefont
  {{Wilson}}},\ }\href {https://doi.org/10.1086/163343} {\bibfield  {journal}
  {\bibinfo  {journal} {\apj}\ }\textbf {\bibinfo {volume} {295}},\ \bibinfo
  {pages} {14} (\bibinfo {year} {1985})}\BibitemShut {NoStop}%
\bibitem [{\citenamefont {{Woosley}}\ and\ \citenamefont
  {{Hoffman}}(1992)}]{Woos92}%
  \BibitemOpen
  \bibfield  {author} {\bibinfo {author} {\bibfnamefont {S.~E.}\ \bibnamefont
  {{Woosley}}}\ and\ \bibinfo {author} {\bibfnamefont {R.~D.}\ \bibnamefont
  {{Hoffman}}},\ }\href {https://doi.org/10.1086/171644} {\bibfield  {journal}
  {\bibinfo  {journal} {\apj}\ }\textbf {\bibinfo {volume} {395}},\ \bibinfo
  {pages} {202} (\bibinfo {year} {1992})}\BibitemShut {NoStop}%
\bibitem [{\citenamefont {Bliss}\ \emph {et~al.}(2018)\citenamefont {Bliss},
  \citenamefont {Witt}, \citenamefont {Arcones}, \citenamefont {Montes},\ and\
  \citenamefont {Pereira}}]{Bliss_2018}%
  \BibitemOpen
  \bibfield  {author} {\bibinfo {author} {\bibfnamefont {J.}~\bibnamefont
  {Bliss}}, \bibinfo {author} {\bibfnamefont {M.}~\bibnamefont {Witt}},
  \bibinfo {author} {\bibfnamefont {A.}~\bibnamefont {Arcones}}, \bibinfo
  {author} {\bibfnamefont {F.}~\bibnamefont {Montes}},\ and\ \bibinfo {author}
  {\bibfnamefont {J.}~\bibnamefont {Pereira}},\ }\href
  {https://doi.org/10.3847/1538-4357/aaadbe} {\bibfield  {journal} {\bibinfo
  {journal} {\apj}\ }\textbf {\bibinfo {volume} {855}},\ \bibinfo {pages} {135}
  (\bibinfo {year} {2018})}\BibitemShut {NoStop}%
\bibitem [{\citenamefont {Bliss}\ \emph {et~al.}(2017)\citenamefont {Bliss},
  \citenamefont {Arcones}, \citenamefont {Montes},\ and\ \citenamefont
  {Pereira}}]{Bliss_2017}%
  \BibitemOpen
  \bibfield  {author} {\bibinfo {author} {\bibfnamefont {J.}~\bibnamefont
  {Bliss}}, \bibinfo {author} {\bibfnamefont {A.}~\bibnamefont {Arcones}},
  \bibinfo {author} {\bibfnamefont {F.}~\bibnamefont {Montes}},\ and\ \bibinfo
  {author} {\bibfnamefont {J.}~\bibnamefont {Pereira}},\ }\href
  {https://doi.org/10.1088/1361-6471/aa63bd} {\bibfield  {journal} {\bibinfo
  {journal} {J. Phys. G}\ }\textbf {\bibinfo {volume} {44}},\ \bibinfo {pages}
  {054003} (\bibinfo {year} {2017})}\BibitemShut {NoStop}%
\bibitem [{\citenamefont {{Chowdhury}}\ \emph {et~al.}(1995)\citenamefont
  {{Chowdhury}}, \citenamefont {{Pal}}, \citenamefont {{Saha}},\ and\
  \citenamefont {{Gangadharan}}}]{Chow95}%
  \BibitemOpen
  \bibfield  {author} {\bibinfo {author} {\bibfnamefont {D.~P.}\ \bibnamefont
  {{Chowdhury}}}, \bibinfo {author} {\bibfnamefont {S.}~\bibnamefont {{Pal}}},
  \bibinfo {author} {\bibfnamefont {S.~K.}\ \bibnamefont {{Saha}}},\ and\
  \bibinfo {author} {\bibfnamefont {S.}~\bibnamefont {{Gangadharan}}},\ }\href
  {https://doi.org/10.1016/0168-583X(95)00663-X} {\bibfield  {journal}
  {\bibinfo  {journal} {Nucl. Instrum. Meth. Phys. Res. Sec. B}\ }\textbf
  {\bibinfo {volume} {103}},\ \bibinfo {pages} {261} (\bibinfo {year}
  {1995})}\BibitemShut {NoStop}%
\bibitem [{\citenamefont {Villa}\ \emph {et~al.}(2020)\citenamefont {Villa},
  \citenamefont {Skuridin}, \citenamefont {Golovkov},\ and\ \citenamefont
  {Garapatsky}}]{Vill20}%
  \BibitemOpen
  \bibfield  {author} {\bibinfo {author} {\bibfnamefont {N.}~\bibnamefont
  {Villa}}, \bibinfo {author} {\bibfnamefont {V.}~\bibnamefont {Skuridin}},
  \bibinfo {author} {\bibfnamefont {V.}~\bibnamefont {Golovkov}},\ and\
  \bibinfo {author} {\bibfnamefont {A.}~\bibnamefont {Garapatsky}},\ }\href
  {https://doi.org/10.1016/j.apradiso.2020.109367} {\bibfield  {journal}
  {\bibinfo  {journal} {Appl. Radiat. Isotopes}\ }\textbf {\bibinfo {volume}
  {166}},\ \bibinfo {pages} {109367} (\bibinfo {year} {2020})}\BibitemShut
  {NoStop}%
\bibitem [{\citenamefont {{Kiss}}\ \emph {et~al.}(2021)\citenamefont {{Kiss}},
  \citenamefont {{Szegedi}}, \citenamefont {{Mohr}}, \citenamefont {{Jacobi}},
  \citenamefont {{Gy{\"u}rky}}, \citenamefont {{Husz{\'a}nk}},\ and\
  \citenamefont {{Arcones}}}]{Kiss21}%
  \BibitemOpen
  \bibfield  {author} {\bibinfo {author} {\bibfnamefont {G.~G.}\ \bibnamefont
  {{Kiss}}}, \bibinfo {author} {\bibfnamefont {T.~N.}\ \bibnamefont
  {{Szegedi}}}, \bibinfo {author} {\bibfnamefont {P.}~\bibnamefont {{Mohr}}},
  \bibinfo {author} {\bibfnamefont {M.}~\bibnamefont {{Jacobi}}}, \bibinfo
  {author} {\bibfnamefont {G.}~\bibnamefont {{Gy{\"u}rky}}}, \bibinfo {author}
  {\bibfnamefont {R.}~\bibnamefont {{Husz{\'a}nk}}},\ and\ \bibinfo {author}
  {\bibfnamefont {A.}~\bibnamefont {{Arcones}}},\ }\href
  {https://doi.org/10.3847/1538-4357/abd2bc} {\bibfield  {journal} {\bibinfo
  {journal} {\apj}\ }\textbf {\bibinfo {volume} {908}},\ \bibinfo {eid} {202}
  (\bibinfo {year} {2021})}\BibitemShut {NoStop}%
\bibitem [{\citenamefont {Boschi}\ \emph {et~al.}(2019)\citenamefont {Boschi},
  \citenamefont {Uccelli},\ and\ \citenamefont {Martini}}]{Bosc19}%
  \BibitemOpen
  \bibfield  {author} {\bibinfo {author} {\bibfnamefont {A.}~\bibnamefont
  {Boschi}}, \bibinfo {author} {\bibfnamefont {L.}~\bibnamefont {Uccelli}},\
  and\ \bibinfo {author} {\bibfnamefont {P.}~\bibnamefont {Martini}},\ }\href
  {https://doi.org/10.3390/app9122526} {\bibfield  {journal} {\bibinfo
  {journal} {Appl. Sci.}\ }\textbf {\bibinfo {volume} {9}},\ \bibinfo {pages}
  {2526} (\bibinfo {year} {2019})}\BibitemShut {NoStop}%
\bibitem [{IAE(2013)}]{IAEA2013}%
  \BibitemOpen
  \bibinfo {title} {Non-heu production technologies for molybdenum-99 and
  technetium-99m}\ (\bibinfo  {publisher} {INTERNATIONAL ATOMIC ENERGY
  AGENCY},\ \bibinfo {address} {Vienna},\ \bibinfo {year} {2013})\BibitemShut
  {NoStop}%
\bibitem [{\citenamefont {Pupillo}\ \emph {et~al.}(2015)\citenamefont
  {Pupillo}, \citenamefont {Esposito}, \citenamefont {Haddad}, \citenamefont
  {Michel},\ and\ \citenamefont {Gambaccini}}]{Pupi15}%
  \BibitemOpen
  \bibfield  {author} {\bibinfo {author} {\bibfnamefont {G.}~\bibnamefont
  {Pupillo}}, \bibinfo {author} {\bibfnamefont {J.}~\bibnamefont {Esposito}},
  \bibinfo {author} {\bibfnamefont {F.}~\bibnamefont {Haddad}}, \bibinfo
  {author} {\bibfnamefont {N.}~\bibnamefont {Michel}},\ and\ \bibinfo {author}
  {\bibfnamefont {M.}~\bibnamefont {Gambaccini}},\ }\href
  {https://doi.org/10.1007/s10967-015-4091-8} {\bibfield  {journal} {\bibinfo
  {journal} {J. Rad. Nuc. Chem.}\ }\textbf {\bibinfo {volume} {305}},\ \bibinfo
  {pages} {73} (\bibinfo {year} {2015})}\BibitemShut {NoStop}%
\bibitem [{\citenamefont {Pupillo}\ \emph {et~al.}(2014)\citenamefont
  {Pupillo}, \citenamefont {Esposito}, \citenamefont {Gambaccini},
  \citenamefont {Haddad},\ and\ \citenamefont {Michel}}]{Pupi14}%
  \BibitemOpen
  \bibfield  {author} {\bibinfo {author} {\bibfnamefont {G.}~\bibnamefont
  {Pupillo}}, \bibinfo {author} {\bibfnamefont {J.}~\bibnamefont {Esposito}},
  \bibinfo {author} {\bibfnamefont {M.}~\bibnamefont {Gambaccini}}, \bibinfo
  {author} {\bibfnamefont {F.}~\bibnamefont {Haddad}},\ and\ \bibinfo {author}
  {\bibfnamefont {N.}~\bibnamefont {Michel}},\ }\href
  {https://doi.org/10.1007/s10967-014-3321-9} {\bibfield  {journal} {\bibinfo
  {journal} {J. Rad. Nuc. Chem.}\ }\textbf {\bibinfo {volume} {302}},\ \bibinfo
  {pages} {911} (\bibinfo {year} {2014})}\BibitemShut {NoStop}%
\bibitem [{\citenamefont {Murata}\ \emph {et~al.}(2019)\citenamefont {Murata},
  \citenamefont {Aikawa}, \citenamefont {Saito}, \citenamefont {Ukon},
  \citenamefont {Komori}, \citenamefont {Haba},\ and\ \citenamefont
  {Tak\'{a}cs}}]{Mura19}%
  \BibitemOpen
  \bibfield  {author} {\bibinfo {author} {\bibfnamefont {T.}~\bibnamefont
  {Murata}}, \bibinfo {author} {\bibfnamefont {M.}~\bibnamefont {Aikawa}},
  \bibinfo {author} {\bibfnamefont {M.}~\bibnamefont {Saito}}, \bibinfo
  {author} {\bibfnamefont {N.}~\bibnamefont {Ukon}}, \bibinfo {author}
  {\bibfnamefont {Y.}~\bibnamefont {Komori}}, \bibinfo {author} {\bibfnamefont
  {H.}~\bibnamefont {Haba}},\ and\ \bibinfo {author} {\bibfnamefont
  {S.}~\bibnamefont {Tak\'{a}cs}},\ }\href
  {https://doi.org/10.1016/j.apradiso.2018.11.012} {\bibfield  {journal}
  {\bibinfo  {journal} {Appl. Radiat. Isotopes}\ }\textbf {\bibinfo {volume}
  {144}},\ \bibinfo {pages} {47} (\bibinfo {year} {2019})}\BibitemShut
  {NoStop}%
\bibitem [{\citenamefont {{Meisel}}\ \emph {et~al.}(2017)\citenamefont
  {{Meisel}}, \citenamefont {{Brune}}, \citenamefont {{Grimes}}, \citenamefont
  {{Ingram}}, \citenamefont {{Massey}},\ and\ \citenamefont
  {{Voinov}}}]{Meis17}%
  \BibitemOpen
  \bibfield  {author} {\bibinfo {author} {\bibfnamefont {Z.}~\bibnamefont
  {{Meisel}}}, \bibinfo {author} {\bibfnamefont {C.~R.}\ \bibnamefont
  {{Brune}}}, \bibinfo {author} {\bibfnamefont {S.~M.}\ \bibnamefont
  {{Grimes}}}, \bibinfo {author} {\bibfnamefont {D.~C.}\ \bibnamefont
  {{Ingram}}}, \bibinfo {author} {\bibfnamefont {T.~N.}\ \bibnamefont
  {{Massey}}},\ and\ \bibinfo {author} {\bibfnamefont {A.~V.}\ \bibnamefont
  {{Voinov}}},\ }\href {https://doi.org/10.1016/j.phpro.2017.09.050} {\bibfield
   {journal} {\bibinfo  {journal} {Physcs. Proc.}\ }\textbf {\bibinfo {volume}
  {90}},\ \bibinfo {pages} {448} (\bibinfo {year} {2017})}\BibitemShut
  {NoStop}%
\bibitem [{\citenamefont {{Browne}}\ and\ \citenamefont
  {{Tuli}}(2017)}]{Brow17}%
  \BibitemOpen
  \bibfield  {author} {\bibinfo {author} {\bibfnamefont {E.}~\bibnamefont
  {{Browne}}}\ and\ \bibinfo {author} {\bibfnamefont {J.~K.}\ \bibnamefont
  {{Tuli}}},\ }\href {https://doi.org/10.1016/j.nds.2017.09.002} {\bibfield
  {journal} {\bibinfo  {journal} {Nucl. Data Sheets}\ }\textbf {\bibinfo
  {volume} {145}},\ \bibinfo {pages} {25} (\bibinfo {year} {2017})}\BibitemShut
  {NoStop}%
\bibitem [{\citenamefont {Goswamy}\ \emph {et~al.}(1992)\citenamefont
  {Goswamy}, \citenamefont {Chand}, \citenamefont {Mehta}, \citenamefont
  {Singh},\ and\ \citenamefont {Trehan}}]{Gosw92}%
  \BibitemOpen
  \bibfield  {author} {\bibinfo {author} {\bibfnamefont {J.}~\bibnamefont
  {Goswamy}}, \bibinfo {author} {\bibfnamefont {B.}~\bibnamefont {Chand}},
  \bibinfo {author} {\bibfnamefont {D.}~\bibnamefont {Mehta}}, \bibinfo
  {author} {\bibfnamefont {N.}~\bibnamefont {Singh}},\ and\ \bibinfo {author}
  {\bibfnamefont {P.}~\bibnamefont {Trehan}},\ }\href
  {https://doi.org/10.1016/0883-2889(92)90173-C} {\bibfield  {journal}
  {\bibinfo  {journal} {International Journal of Radiation Applications and
  Instrumentation. Part A. Appl. Radiat. Isotopes}\ }\textbf {\bibinfo {volume}
  {43}},\ \bibinfo {pages} {1467} (\bibinfo {year} {1992})}\BibitemShut
  {NoStop}%
\bibitem [{\citenamefont {Semkow}\ \emph {et~al.}(1990)\citenamefont {Semkow},
  \citenamefont {Mehmood}, \citenamefont {Parekh},\ and\ \citenamefont
  {Virgil}}]{SEMKOW1990437}%
  \BibitemOpen
  \bibfield  {author} {\bibinfo {author} {\bibfnamefont {T.~M.}\ \bibnamefont
  {Semkow}}, \bibinfo {author} {\bibfnamefont {G.}~\bibnamefont {Mehmood}},
  \bibinfo {author} {\bibfnamefont {P.~P.}\ \bibnamefont {Parekh}},\ and\
  \bibinfo {author} {\bibfnamefont {M.}~\bibnamefont {Virgil}},\ }\href
  {https://doi.org/https://doi.org/10.1016/0168-9002(90)90561-J} {\bibfield
  {journal} {\bibinfo  {journal} {Nucl. Instrum. Meth. Phys. Res. Sec. A}\
  }\textbf {\bibinfo {volume} {290}},\ \bibinfo {pages} {437 } (\bibinfo {year}
  {1990})}\BibitemShut {NoStop}%
\bibitem [{\citenamefont {Knoll}(2010)}]{knoll}%
  \BibitemOpen
  \bibfield  {author} {\bibinfo {author} {\bibfnamefont {G.~F.}\ \bibnamefont
  {Knoll}},\ }\bibinfo {title} {Radiation detection and measurement, 4$^{th}$
  edition}\ (\bibinfo  {publisher} {Wiley},\ \bibinfo {address} {Dordrecht},\
  \bibinfo {year} {2010})\ p.\ \bibinfo {pages} {857}\BibitemShut {NoStop}%
\bibitem [{\citenamefont {{Ziegler}}\ \emph {et~al.}(2010)\citenamefont
  {{Ziegler}}, \citenamefont {{Ziegler}},\ and\ \citenamefont
  {{Biersack}}}]{Zieg10}%
  \BibitemOpen
  \bibfield  {author} {\bibinfo {author} {\bibfnamefont {J.~F.}\ \bibnamefont
  {{Ziegler}}}, \bibinfo {author} {\bibfnamefont {M.~D.}\ \bibnamefont
  {{Ziegler}}},\ and\ \bibinfo {author} {\bibfnamefont {J.~P.}\ \bibnamefont
  {{Biersack}}},\ }\href {https://doi.org/10.1016/j.nimb.2010.02.091}
  {\bibfield  {journal} {\bibinfo  {journal} {Nucl. Instrum. Meth. Phys. Res.
  Sec. B}\ }\textbf {\bibinfo {volume} {268}},\ \bibinfo {pages} {1818}
  (\bibinfo {year} {2010})}\BibitemShut {NoStop}%
\bibitem [{\citenamefont {{Paul}}\ and\ \citenamefont
  {{Schinner}}(2005)}]{Paul05}%
  \BibitemOpen
  \bibfield  {author} {\bibinfo {author} {\bibfnamefont {H.}~\bibnamefont
  {{Paul}}}\ and\ \bibinfo {author} {\bibfnamefont {A.}~\bibnamefont
  {{Schinner}}},\ }\href {https://doi.org/10.1016/j.nimb.2004.10.007}
  {\bibfield  {journal} {\bibinfo  {journal} {Nucl. Instrum. Meth. Phys. Res.
  Sec. B}\ }\textbf {\bibinfo {volume} {227}},\ \bibinfo {pages} {461}
  (\bibinfo {year} {2005})}\BibitemShut {NoStop}%
\bibitem [{\citenamefont {{Lin}}\ \emph {et~al.}(1973)\citenamefont {{Lin}},
  \citenamefont {{Olson}},\ and\ \citenamefont {{Powers}}}]{Lin73}%
  \BibitemOpen
  \bibfield  {author} {\bibinfo {author} {\bibfnamefont {W.~K.}\ \bibnamefont
  {{Lin}}}, \bibinfo {author} {\bibfnamefont {H.~G.}\ \bibnamefont {{Olson}}},\
  and\ \bibinfo {author} {\bibfnamefont {D.}~\bibnamefont {{Powers}}},\ }\href
  {https://doi.org/10.1103/PhysRevB.8.1881} {\bibfield  {journal} {\bibinfo
  {journal} {\prb}\ }\textbf {\bibinfo {volume} {8}},\ \bibinfo {pages} {1881}
  (\bibinfo {year} {1973})}\BibitemShut {NoStop}%
\bibitem [{\citenamefont {Hauser}\ and\ \citenamefont
  {Feshbach}(1952)}]{Hauser1}%
  \BibitemOpen
  \bibfield  {author} {\bibinfo {author} {\bibfnamefont {W.}~\bibnamefont
  {Hauser}}\ and\ \bibinfo {author} {\bibfnamefont {H.}~\bibnamefont
  {Feshbach}},\ }\href {https://doi.org/10.1103/PhysRev.87.366} {\bibfield
  {journal} {\bibinfo  {journal} {Phys. Rev.}\ }\textbf {\bibinfo {volume}
  {87}},\ \bibinfo {pages} {366} (\bibinfo {year} {1952})}\BibitemShut
  {NoStop}%
\bibitem [{\citenamefont {{Wolfenstein}}(1951)}]{Wolf51}%
  \BibitemOpen
  \bibfield  {author} {\bibinfo {author} {\bibfnamefont {L.}~\bibnamefont
  {{Wolfenstein}}},\ }\href {https://doi.org/10.1103/PhysRev.82.690} {\bibfield
   {journal} {\bibinfo  {journal} {Phys. Rev.}\ }\textbf {\bibinfo {volume}
  {82}},\ \bibinfo {pages} {690} (\bibinfo {year} {1951})}\BibitemShut
  {NoStop}%
\bibitem [{Kon()}]{Koni08}%
  \BibitemOpen
  \href@noop {} {}\bibinfo {note} {A.J. Koning, S. Hilaire and M.C.
  Duijvestijn, ``TALYS-1.0", Proceedings of the International Conference on
  Nuclear Data for Science and Technology - ND2007, April 22-27, 2007, Nice,
  France, eds. O. Bersillon, F. Gunsing, E. Bauge, R. Jacqmin and S. Leray, EDP
  Sciences, 2008, p. 211-214}\BibitemShut {NoStop}%
\bibitem [{\citenamefont {Lahanas}\ \emph {et~al.}(1986)\citenamefont
  {Lahanas}, \citenamefont {Rychel}, \citenamefont {Singh}, \citenamefont
  {Gyufko}, \citenamefont {Kolbert}, \citenamefont {{Van Kr\"{u}chten}},
  \citenamefont {Madadakis},\ and\ \citenamefont {Wiedner}}]{Laha86}%
  \BibitemOpen
  \bibfield  {author} {\bibinfo {author} {\bibfnamefont {M.}~\bibnamefont
  {Lahanas}}, \bibinfo {author} {\bibfnamefont {D.}~\bibnamefont {Rychel}},
  \bibinfo {author} {\bibfnamefont {P.}~\bibnamefont {Singh}}, \bibinfo
  {author} {\bibfnamefont {R.}~\bibnamefont {Gyufko}}, \bibinfo {author}
  {\bibfnamefont {D.}~\bibnamefont {Kolbert}}, \bibinfo {author} {\bibfnamefont
  {B.}~\bibnamefont {{Van Kr\"{u}chten}}}, \bibinfo {author} {\bibfnamefont
  {E.}~\bibnamefont {Madadakis}},\ and\ \bibinfo {author} {\bibfnamefont
  {C.}~\bibnamefont {Wiedner}},\ }\href
  {https://doi.org/https://doi.org/10.1016/0375-9474(86)90314-3} {\bibfield
  {journal} {\bibinfo  {journal} {Nucl. Phys. A}\ }\textbf {\bibinfo {volume}
  {455}},\ \bibinfo {pages} {399} (\bibinfo {year} {1986})}\BibitemShut
  {NoStop}%
\bibitem [{\citenamefont {Lund}\ \emph {et~al.}(1995)\citenamefont {Lund},
  \citenamefont {Bateman}, \citenamefont {Utku}, \citenamefont {Horen},\ and\
  \citenamefont {Satchler}}]{Lund95}%
  \BibitemOpen
  \bibfield  {author} {\bibinfo {author} {\bibfnamefont {B.~J.}\ \bibnamefont
  {Lund}}, \bibinfo {author} {\bibfnamefont {N.~P.~T.}\ \bibnamefont
  {Bateman}}, \bibinfo {author} {\bibfnamefont {S.}~\bibnamefont {Utku}},
  \bibinfo {author} {\bibfnamefont {D.~J.}\ \bibnamefont {Horen}},\ and\
  \bibinfo {author} {\bibfnamefont {G.~R.}\ \bibnamefont {Satchler}},\ }\href
  {https://doi.org/10.1103/PhysRevC.51.635} {\bibfield  {journal} {\bibinfo
  {journal} {Phys. Rev. C}\ }\textbf {\bibinfo {volume} {51}},\ \bibinfo
  {pages} {635} (\bibinfo {year} {1995})}\BibitemShut {NoStop}%
\bibitem [{\citenamefont {{Larsen}}\ \emph {et~al.}(2019)\citenamefont
  {{Larsen}}, \citenamefont {{Spyrou}}, \citenamefont {{Liddick}},\ and\
  \citenamefont {{Guttormsen}}}]{Lars19}%
  \BibitemOpen
  \bibfield  {author} {\bibinfo {author} {\bibfnamefont {A.~C.}\ \bibnamefont
  {{Larsen}}}, \bibinfo {author} {\bibfnamefont {A.}~\bibnamefont {{Spyrou}}},
  \bibinfo {author} {\bibfnamefont {S.~N.}\ \bibnamefont {{Liddick}}},\ and\
  \bibinfo {author} {\bibfnamefont {M.}~\bibnamefont {{Guttormsen}}},\ }\href
  {https://doi.org/10.1016/j.ppnp.2019.04.002} {\bibfield  {journal} {\bibinfo
  {journal} {Prog. Part. Nucl. Phys.}\ }\textbf {\bibinfo {volume} {107}},\
  \bibinfo {pages} {69} (\bibinfo {year} {2019})}\BibitemShut {NoStop}%
\bibitem [{\citenamefont {{Mohr}}(2016)}]{Mohr16}%
  \BibitemOpen
  \bibfield  {author} {\bibinfo {author} {\bibfnamefont {P.}~\bibnamefont
  {{Mohr}}},\ }\href {https://doi.org/10.1103/PhysRevC.94.035801} {\bibfield
  {journal} {\bibinfo  {journal} {\prc}\ }\textbf {\bibinfo {volume} {94}},\
  \bibinfo {eid} {035801} (\bibinfo {year} {2016})}\BibitemShut {NoStop}%
\bibitem [{\citenamefont {McFadden}\ and\ \citenamefont
  {Satchler}(1966)}]{mcfadden}%
  \BibitemOpen
  \bibfield  {author} {\bibinfo {author} {\bibfnamefont {L.}~\bibnamefont
  {McFadden}}\ and\ \bibinfo {author} {\bibfnamefont {G.}~\bibnamefont
  {Satchler}},\ }\href@noop {} {\bibfield  {journal} {\bibinfo  {journal}
  {Nucl. Phys.}\ }\textbf {\bibinfo {volume} {84}},\ \bibinfo {pages} {177}
  (\bibinfo {year} {1966})}\BibitemShut {NoStop}%
\bibitem [{\citenamefont {{Mohr}}(2015)}]{Mohr15}%
  \BibitemOpen
  \bibfield  {author} {\bibinfo {author} {\bibfnamefont {P.}~\bibnamefont
  {{Mohr}}},\ }\href {https://doi.org/10.1140/epja/i2015-15056-5} {\bibfield
  {journal} {\bibinfo  {journal} {European Physical Journal A}\ }\textbf
  {\bibinfo {volume} {51}},\ \bibinfo {eid} {56} (\bibinfo {year}
  {2015})}\BibitemShut {NoStop}%
\bibitem [{\citenamefont {Dilg}\ \emph {et~al.}(1973)\citenamefont {Dilg},
  \citenamefont {Schantl}, \citenamefont {Vonach},\ and\ \citenamefont
  {Uhl}}]{Dilg73}%
  \BibitemOpen
  \bibfield  {author} {\bibinfo {author} {\bibfnamefont {W.}~\bibnamefont
  {Dilg}}, \bibinfo {author} {\bibfnamefont {W.}~\bibnamefont {Schantl}},
  \bibinfo {author} {\bibfnamefont {H.}~\bibnamefont {Vonach}},\ and\ \bibinfo
  {author} {\bibfnamefont {M.}~\bibnamefont {Uhl}},\ }\href
  {https://doi.org/https://doi.org/10.1016/0375-9474(73)90196-6} {\bibfield
  {journal} {\bibinfo  {journal} {Nucl. Phys. A}\ }\textbf {\bibinfo {volume}
  {217}},\ \bibinfo {pages} {269 } (\bibinfo {year} {1973})}\BibitemShut
  {NoStop}%
\bibitem [{\citenamefont {{Chankova}}\ \emph {et~al.}(2006)\citenamefont
  {{Chankova}}, \citenamefont {{Schiller}}, \citenamefont {{Agvaanluvsan}},
  \citenamefont {{Algin}}, \citenamefont {{Bernstein}}, \citenamefont
  {{Guttormsen}}, \citenamefont {{Ingebretsen}}, \citenamefont
  {{L{\"o}nnroth}}, \citenamefont {{Messelt}}, \citenamefont {{Mitchell}},
  \citenamefont {{Rekstad}}, \citenamefont {{Siem}}, \citenamefont {{Larsen}},
  \citenamefont {{Voinov}},\ and\ \citenamefont
  {{{\O}deg{\r{a}}rd}}}]{Chank06}%
  \BibitemOpen
  \bibfield  {author} {\bibinfo {author} {\bibfnamefont {R.}~\bibnamefont
  {{Chankova}}}, \bibinfo {author} {\bibfnamefont {A.}~\bibnamefont
  {{Schiller}}}, \bibinfo {author} {\bibfnamefont {U.}~\bibnamefont
  {{Agvaanluvsan}}}, \bibinfo {author} {\bibfnamefont {E.}~\bibnamefont
  {{Algin}}}, \bibinfo {author} {\bibfnamefont {L.~A.}\ \bibnamefont
  {{Bernstein}}}, \bibinfo {author} {\bibfnamefont {M.}~\bibnamefont
  {{Guttormsen}}}, \bibinfo {author} {\bibfnamefont {F.}~\bibnamefont
  {{Ingebretsen}}}, \bibinfo {author} {\bibfnamefont {T.}~\bibnamefont
  {{L{\"o}nnroth}}}, \bibinfo {author} {\bibfnamefont {S.}~\bibnamefont
  {{Messelt}}}, \bibinfo {author} {\bibfnamefont {G.~E.}\ \bibnamefont
  {{Mitchell}}}, \bibinfo {author} {\bibfnamefont {J.}~\bibnamefont
  {{Rekstad}}}, \bibinfo {author} {\bibfnamefont {S.}~\bibnamefont {{Siem}}},
  \bibinfo {author} {\bibfnamefont {A.~C.}\ \bibnamefont {{Larsen}}}, \bibinfo
  {author} {\bibfnamefont {A.}~\bibnamefont {{Voinov}}},\ and\ \bibinfo
  {author} {\bibfnamefont {S.}~\bibnamefont {{{\O}deg{\r{a}}rd}}},\ }\href
  {https://doi.org/10.1103/PhysRevC.73.034311} {\bibfield  {journal} {\bibinfo
  {journal} {\prc}\ }\textbf {\bibinfo {volume} {73}},\ \bibinfo {eid} {034311}
  (\bibinfo {year} {2006})}\BibitemShut {NoStop}%
\bibitem [{\citenamefont {{Martin}}\ \emph {et~al.}(2017)\citenamefont
  {{Martin}}, \citenamefont {{von Neumann-Cosel}}, \citenamefont {{Tamii}},
  \citenamefont {{Aoi}}, \citenamefont {{Bassauer}}, \citenamefont
  {{Bertulani}}, \citenamefont {{Carter}}, \citenamefont {{Donaldson}},
  \citenamefont {{Fujita}}, \citenamefont {{Fujita}}, \citenamefont
  {{Hashimoto}}, \citenamefont {{Hatanaka}}, \citenamefont {{Ito}},
  \citenamefont {{Krugmann}}, \citenamefont {{Liu}}, \citenamefont {{Maeda}},
  \citenamefont {{Miki}}, \citenamefont {{Neveling}}, \citenamefont
  {{Pietralla}}, \citenamefont {{Poltoratska}}, \citenamefont {{Ponomarev}},
  \citenamefont {{Richter}}, \citenamefont {{Shima}}, \citenamefont
  {{Yamamoto}},\ and\ \citenamefont {{Zweidinger}}}]{Mart17}%
  \BibitemOpen
  \bibfield  {author} {\bibinfo {author} {\bibfnamefont {D.}~\bibnamefont
  {{Martin}}}, \bibinfo {author} {\bibfnamefont {P.}~\bibnamefont {{von
  Neumann-Cosel}}}, \bibinfo {author} {\bibfnamefont {A.}~\bibnamefont
  {{Tamii}}}, \bibinfo {author} {\bibfnamefont {N.}~\bibnamefont {{Aoi}}},
  \bibinfo {author} {\bibfnamefont {S.}~\bibnamefont {{Bassauer}}}, \bibinfo
  {author} {\bibfnamefont {C.~A.}\ \bibnamefont {{Bertulani}}}, \bibinfo
  {author} {\bibfnamefont {J.}~\bibnamefont {{Carter}}}, \bibinfo {author}
  {\bibfnamefont {L.}~\bibnamefont {{Donaldson}}}, \bibinfo {author}
  {\bibfnamefont {H.}~\bibnamefont {{Fujita}}}, \bibinfo {author}
  {\bibfnamefont {Y.}~\bibnamefont {{Fujita}}}, \bibinfo {author}
  {\bibfnamefont {T.}~\bibnamefont {{Hashimoto}}}, \bibinfo {author}
  {\bibfnamefont {K.}~\bibnamefont {{Hatanaka}}}, \bibinfo {author}
  {\bibfnamefont {T.}~\bibnamefont {{Ito}}}, \bibinfo {author} {\bibfnamefont
  {A.}~\bibnamefont {{Krugmann}}}, \bibinfo {author} {\bibfnamefont
  {B.}~\bibnamefont {{Liu}}}, \bibinfo {author} {\bibfnamefont
  {Y.}~\bibnamefont {{Maeda}}}, \bibinfo {author} {\bibfnamefont
  {K.}~\bibnamefont {{Miki}}}, \bibinfo {author} {\bibfnamefont
  {R.}~\bibnamefont {{Neveling}}}, \bibinfo {author} {\bibfnamefont
  {N.}~\bibnamefont {{Pietralla}}}, \bibinfo {author} {\bibfnamefont
  {I.}~\bibnamefont {{Poltoratska}}}, \bibinfo {author} {\bibfnamefont {V.~Y.}\
  \bibnamefont {{Ponomarev}}}, \bibinfo {author} {\bibfnamefont
  {A.}~\bibnamefont {{Richter}}}, \bibinfo {author} {\bibfnamefont
  {T.}~\bibnamefont {{Shima}}}, \bibinfo {author} {\bibfnamefont
  {T.}~\bibnamefont {{Yamamoto}}},\ and\ \bibinfo {author} {\bibfnamefont
  {M.}~\bibnamefont {{Zweidinger}}},\ }\href
  {https://doi.org/10.1103/PhysRevLett.119.182503} {\bibfield  {journal}
  {\bibinfo  {journal} {\prl}\ }\textbf {\bibinfo {volume} {119}},\ \bibinfo
  {eid} {182503} (\bibinfo {year} {2017})}\BibitemShut {NoStop}%
\bibitem [{\citenamefont {Ignatyuk}\ \emph {et~al.}(1975)\citenamefont
  {Ignatyuk}, \citenamefont {Smirenkin},\ and\ \citenamefont
  {Tishin}}]{Igna75}%
  \BibitemOpen
  \bibfield  {author} {\bibinfo {author} {\bibfnamefont {A.~V.}\ \bibnamefont
  {Ignatyuk}}, \bibinfo {author} {\bibfnamefont {G.~N.}\ \bibnamefont
  {Smirenkin}},\ and\ \bibinfo {author} {\bibfnamefont {A.~S.}\ \bibnamefont
  {Tishin}},\ }\href@noop {} {\bibfield  {journal} {\bibinfo  {journal} {Sov.
  J. Nucl. Phys.}\ }\textbf {\bibinfo {volume} {21}},\ \bibinfo {pages} {255}
  (\bibinfo {year} {1975})}\BibitemShut {NoStop}%
\bibitem [{\citenamefont {{Bethe}}(1936)}]{Beth36}%
  \BibitemOpen
  \bibfield  {author} {\bibinfo {author} {\bibfnamefont {H.~A.}\ \bibnamefont
  {{Bethe}}},\ }\href {https://doi.org/10.1103/PhysRev.50.332} {\bibfield
  {journal} {\bibinfo  {journal} {Phys. Rev.}\ }\textbf {\bibinfo {volume}
  {50}},\ \bibinfo {pages} {332} (\bibinfo {year} {1936})}\BibitemShut
  {NoStop}%
\bibitem [{\citenamefont {{Capote}}\ \emph {et~al.}(2009)\citenamefont
  {{Capote}}, \citenamefont {{Herman}}, \citenamefont {{Oblo{\v{z}}insk{\'y}}},
  \citenamefont {{Young}}, \citenamefont {{Goriely}}, \citenamefont {{Belgya}},
  \citenamefont {{Ignatyuk}}, \citenamefont {{Koning}}, \citenamefont
  {{Hilaire}}, \citenamefont {{Plujko}}, \citenamefont {{Avrigeanu}},
  \citenamefont {{Bersillon}}, \citenamefont {{Chadwick}}, \citenamefont
  {{Fukahori}}, \citenamefont {{Ge}}, \citenamefont {{Han}}, \citenamefont
  {{Kailas}}, \citenamefont {{Kopecky}}, \citenamefont {{Maslov}},
  \citenamefont {{Reffo}}, \citenamefont {{Sin}}, \citenamefont
  {{Soukhovitskii}},\ and\ \citenamefont {{Talou}}}]{Capo09}%
  \BibitemOpen
  \bibfield  {author} {\bibinfo {author} {\bibfnamefont {R.}~\bibnamefont
  {{Capote}}}, \bibinfo {author} {\bibfnamefont {M.}~\bibnamefont {{Herman}}},
  \bibinfo {author} {\bibfnamefont {P.}~\bibnamefont {{Oblo{\v{z}}insk{\'y}}}},
  \bibinfo {author} {\bibfnamefont {P.~G.}\ \bibnamefont {{Young}}}, \bibinfo
  {author} {\bibfnamefont {S.}~\bibnamefont {{Goriely}}}, \bibinfo {author}
  {\bibfnamefont {T.}~\bibnamefont {{Belgya}}}, \bibinfo {author}
  {\bibfnamefont {A.~V.}\ \bibnamefont {{Ignatyuk}}}, \bibinfo {author}
  {\bibfnamefont {A.~J.}\ \bibnamefont {{Koning}}}, \bibinfo {author}
  {\bibfnamefont {S.}~\bibnamefont {{Hilaire}}}, \bibinfo {author}
  {\bibfnamefont {V.~A.}\ \bibnamefont {{Plujko}}}, \bibinfo {author}
  {\bibfnamefont {M.}~\bibnamefont {{Avrigeanu}}}, \bibinfo {author}
  {\bibfnamefont {O.}~\bibnamefont {{Bersillon}}}, \bibinfo {author}
  {\bibfnamefont {M.~B.}\ \bibnamefont {{Chadwick}}}, \bibinfo {author}
  {\bibfnamefont {T.}~\bibnamefont {{Fukahori}}}, \bibinfo {author}
  {\bibfnamefont {Z.}~\bibnamefont {{Ge}}}, \bibinfo {author} {\bibfnamefont
  {Y.}~\bibnamefont {{Han}}}, \bibinfo {author} {\bibfnamefont
  {S.}~\bibnamefont {{Kailas}}}, \bibinfo {author} {\bibfnamefont
  {J.}~\bibnamefont {{Kopecky}}}, \bibinfo {author} {\bibfnamefont {V.~M.}\
  \bibnamefont {{Maslov}}}, \bibinfo {author} {\bibfnamefont {G.}~\bibnamefont
  {{Reffo}}}, \bibinfo {author} {\bibfnamefont {M.}~\bibnamefont {{Sin}}},
  \bibinfo {author} {\bibfnamefont {E.~S.}\ \bibnamefont {{Soukhovitskii}}},\
  and\ \bibinfo {author} {\bibfnamefont {P.}~\bibnamefont {{Talou}}},\ }\href
  {https://doi.org/10.1016/j.nds.2009.10.004} {\bibfield  {journal} {\bibinfo
  {journal} {Nucl. Data Sheets}\ }\textbf {\bibinfo {volume} {110}},\ \bibinfo
  {pages} {3107} (\bibinfo {year} {2009})}\BibitemShut {NoStop}%
\bibitem [{\citenamefont {{Grimes}}\ \emph {et~al.}(2016)\citenamefont
  {{Grimes}}, \citenamefont {{Voinov}},\ and\ \citenamefont
  {{Massey}}}]{Grim16}%
  \BibitemOpen
  \bibfield  {author} {\bibinfo {author} {\bibfnamefont {S.~M.}\ \bibnamefont
  {{Grimes}}}, \bibinfo {author} {\bibfnamefont {A.~V.}\ \bibnamefont
  {{Voinov}}},\ and\ \bibinfo {author} {\bibfnamefont {T.~N.}\ \bibnamefont
  {{Massey}}},\ }\href {https://doi.org/10.1103/PhysRevC.94.014308} {\bibfield
  {journal} {\bibinfo  {journal} {\prc}\ }\textbf {\bibinfo {volume} {94}},\
  \bibinfo {eid} {014308} (\bibinfo {year} {2016})}\BibitemShut {NoStop}%
\bibitem [{\citenamefont {Press}\ \emph {et~al.}(1992)\citenamefont {Press},
  \citenamefont {Teukolsky}, \citenamefont {Vetterling},\ and\ \citenamefont
  {Flannery}}]{Pres92}%
  \BibitemOpen
  \bibfield  {author} {\bibinfo {author} {\bibfnamefont {W.~H.}\ \bibnamefont
  {Press}}, \bibinfo {author} {\bibfnamefont {S.~A.}\ \bibnamefont
  {Teukolsky}}, \bibinfo {author} {\bibfnamefont {W.~T.}\ \bibnamefont
  {Vetterling}},\ and\ \bibinfo {author} {\bibfnamefont {B.~P.}\ \bibnamefont
  {Flannery}},\ }\href@noop {} {\emph {\bibinfo {title} {Numerical Recipes in
  {C}: {T}he Art of Scientific Computing}}},\ \bibinfo {edition} {2nd}\ ed.\
  (\bibinfo  {publisher} {Cambridge University, Cambridge, England},\ \bibinfo
  {year} {1992})\BibitemShut {NoStop}%
\bibitem [{\citenamefont {{Szegedi}}\ \emph {et~al.}(2021)\citenamefont
  {{Szegedi}}, \citenamefont {{Kiss}}, \citenamefont {{Mohr}}, \citenamefont
  {{Psaltis}}, \citenamefont {{Jacobi}}, \citenamefont {{Barnaf{\"o}ldi}},
  \citenamefont {{Sz{\"u}cs}}, \citenamefont {{Gy{\"u}rky}},\ and\
  \citenamefont {{Arcones}}}]{Szeg21}%
  \BibitemOpen
  \bibfield  {author} {\bibinfo {author} {\bibfnamefont {T.~N.}\ \bibnamefont
  {{Szegedi}}}, \bibinfo {author} {\bibfnamefont {G.~G.}\ \bibnamefont
  {{Kiss}}}, \bibinfo {author} {\bibfnamefont {P.}~\bibnamefont {{Mohr}}},
  \bibinfo {author} {\bibfnamefont {A.}~\bibnamefont {{Psaltis}}}, \bibinfo
  {author} {\bibfnamefont {M.}~\bibnamefont {{Jacobi}}}, \bibinfo {author}
  {\bibfnamefont {G.~G.}\ \bibnamefont {{Barnaf{\"o}ldi}}}, \bibinfo {author}
  {\bibfnamefont {T.}~\bibnamefont {{Sz{\"u}cs}}}, \bibinfo {author}
  {\bibfnamefont {G.}~\bibnamefont {{Gy{\"u}rky}}},\ and\ \bibinfo {author}
  {\bibfnamefont {A.}~\bibnamefont {{Arcones}}},\ }\href
  {https://doi.org/10.1103/PhysRevC.104.035804} {\bibfield  {journal} {\bibinfo
   {journal} {\prc}\ }\textbf {\bibinfo {volume} {104}},\ \bibinfo {eid}
  {035804} (\bibinfo {year} {2021})}\BibitemShut {NoStop}%
\bibitem [{\citenamefont {{Psaltis}}\ \emph {et~al.}(2022)\citenamefont
  {{Psaltis}}, \citenamefont {{Arcones}}, \citenamefont {{Montes}},
  \citenamefont {{Mohr}}, \citenamefont {{Hansen}}, \citenamefont {{Jacobi}},\
  and\ \citenamefont {{Schatz}}}]{Psal22}%
  \BibitemOpen
  \bibfield  {author} {\bibinfo {author} {\bibfnamefont {A.}~\bibnamefont
  {{Psaltis}}}, \bibinfo {author} {\bibfnamefont {A.}~\bibnamefont
  {{Arcones}}}, \bibinfo {author} {\bibfnamefont {F.}~\bibnamefont {{Montes}}},
  \bibinfo {author} {\bibfnamefont {P.}~\bibnamefont {{Mohr}}}, \bibinfo
  {author} {\bibfnamefont {C.~J.}\ \bibnamefont {{Hansen}}}, \bibinfo {author}
  {\bibfnamefont {M.}~\bibnamefont {{Jacobi}}},\ and\ \bibinfo {author}
  {\bibfnamefont {H.}~\bibnamefont {{Schatz}}},\ }\href@noop {} {\bibfield
  {journal} {\bibinfo  {journal} {arXiv e-prints}\ ,\ \bibinfo {eid}
  {arXiv:2204.07136}} (\bibinfo {year} {2022})},\ \Eprint
  {https://arxiv.org/abs/2204.07136} {arXiv:2204.07136 [astro-ph.HE]}
  \BibitemShut {NoStop}%
\bibitem [{\citenamefont {{Mohr}}\ \emph {et~al.}(2020)\citenamefont {{Mohr}},
  \citenamefont {{F{\"u}l{\"o}p}}, \citenamefont {{Gy{\"u}rky}}, \citenamefont
  {{Kiss}},\ and\ \citenamefont {{Sz{\"u}cs}}}]{Mohr20}%
  \BibitemOpen
  \bibfield  {author} {\bibinfo {author} {\bibfnamefont {P.}~\bibnamefont
  {{Mohr}}}, \bibinfo {author} {\bibfnamefont {Z.}~\bibnamefont
  {{F{\"u}l{\"o}p}}}, \bibinfo {author} {\bibfnamefont {G.}~\bibnamefont
  {{Gy{\"u}rky}}}, \bibinfo {author} {\bibfnamefont {G.~G.}\ \bibnamefont
  {{Kiss}}},\ and\ \bibinfo {author} {\bibfnamefont {T.}~\bibnamefont
  {{Sz{\"u}cs}}},\ }\href {https://doi.org/10.1103/PhysRevLett.124.252701}
  {\bibfield  {journal} {\bibinfo  {journal} {\prl}\ }\textbf {\bibinfo
  {volume} {124}},\ \bibinfo {eid} {252701} (\bibinfo {year}
  {2020})}\BibitemShut {NoStop}%
\bibitem [{\citenamefont {{Nolte}}\ \emph {et~al.}(1987)\citenamefont
  {{Nolte}}, \citenamefont {{Machner}},\ and\ \citenamefont
  {{Bojowald}}}]{Nolt87}%
  \BibitemOpen
  \bibfield  {author} {\bibinfo {author} {\bibfnamefont {M.}~\bibnamefont
  {{Nolte}}}, \bibinfo {author} {\bibfnamefont {H.}~\bibnamefont {{Machner}}},\
  and\ \bibinfo {author} {\bibfnamefont {J.}~\bibnamefont {{Bojowald}}},\
  }\href {https://doi.org/10.1103/PhysRevC.36.1312} {\bibfield  {journal}
  {\bibinfo  {journal} {\prc}\ }\textbf {\bibinfo {volume} {36}},\ \bibinfo
  {pages} {1312} (\bibinfo {year} {1987})}\BibitemShut {NoStop}%
\bibitem [{\citenamefont {{Avrigeanu}}\ \emph {et~al.}(1994)\citenamefont
  {{Avrigeanu}}, \citenamefont {{Hodgson}},\ and\ \citenamefont
  {{Avrigeanu}}}]{Avri94}%
  \BibitemOpen
  \bibfield  {author} {\bibinfo {author} {\bibfnamefont {V.}~\bibnamefont
  {{Avrigeanu}}}, \bibinfo {author} {\bibfnamefont {P.~E.}\ \bibnamefont
  {{Hodgson}}},\ and\ \bibinfo {author} {\bibfnamefont {M.}~\bibnamefont
  {{Avrigeanu}}},\ }\href {https://doi.org/10.1103/PhysRevC.49.2136} {\bibfield
   {journal} {\bibinfo  {journal} {\prc}\ }\textbf {\bibinfo {volume} {49}},\
  \bibinfo {pages} {2136} (\bibinfo {year} {1994})}\BibitemShut {NoStop}%
\bibitem [{\citenamefont {{Avrigeanu}}\ \emph {et~al.}(2014)\citenamefont
  {{Avrigeanu}}, \citenamefont {{Avrigeanu}},\ and\ \citenamefont
  {{M{\v{a}}n{\v{a}}ilescu}}}]{Avri14}%
  \BibitemOpen
  \bibfield  {author} {\bibinfo {author} {\bibfnamefont {V.}~\bibnamefont
  {{Avrigeanu}}}, \bibinfo {author} {\bibfnamefont {M.}~\bibnamefont
  {{Avrigeanu}}},\ and\ \bibinfo {author} {\bibfnamefont {C.}~\bibnamefont
  {{M{\v{a}}n{\v{a}}ilescu}}},\ }\href
  {https://doi.org/10.1103/PhysRevC.90.044612} {\bibfield  {journal} {\bibinfo
  {journal} {\prc}\ }\textbf {\bibinfo {volume} {90}},\ \bibinfo {eid} {044612}
  (\bibinfo {year} {2014})}\BibitemShut {NoStop}%
\bibitem [{\citenamefont {{Rauscher}}\ and\ \citenamefont
  {{Thielemann}}(2000)}]{Raus00}%
  \BibitemOpen
  \bibfield  {author} {\bibinfo {author} {\bibfnamefont {T.}~\bibnamefont
  {{Rauscher}}}\ and\ \bibinfo {author} {\bibfnamefont {F.-K.}\ \bibnamefont
  {{Thielemann}}},\ }\href {https://doi.org/10.1006/adnd.2000.0834} {\bibfield
  {journal} {\bibinfo  {journal} {Atom. Data. Nucl. Data}\ }\textbf {\bibinfo
  {volume} {75}},\ \bibinfo {pages} {1} (\bibinfo {year} {2000})}\BibitemShut
  {NoStop}%
\bibitem [{\citenamefont {{Cyburt}}\ \emph {et~al.}(2010)\citenamefont
  {{Cyburt}}, \citenamefont {{Amthor}}, \citenamefont {{Ferguson}},
  \citenamefont {{Meisel}}, \citenamefont {{Smith}}, \citenamefont {{Warren}},
  \citenamefont {{Heger}}, \citenamefont {{Hoffman}}, \citenamefont
  {{Rauscher}}, \citenamefont {{Sakharuk}}, \citenamefont {{Schatz}},
  \citenamefont {{Thielemann}},\ and\ \citenamefont {{Wiescher}}}]{Cybu10}%
  \BibitemOpen
  \bibfield  {author} {\bibinfo {author} {\bibfnamefont {R.~H.}\ \bibnamefont
  {{Cyburt}}}, \bibinfo {author} {\bibfnamefont {A.~M.}\ \bibnamefont
  {{Amthor}}}, \bibinfo {author} {\bibfnamefont {R.}~\bibnamefont
  {{Ferguson}}}, \bibinfo {author} {\bibfnamefont {Z.}~\bibnamefont
  {{Meisel}}}, \bibinfo {author} {\bibfnamefont {K.}~\bibnamefont {{Smith}}},
  \bibinfo {author} {\bibfnamefont {S.}~\bibnamefont {{Warren}}}, \bibinfo
  {author} {\bibfnamefont {A.}~\bibnamefont {{Heger}}}, \bibinfo {author}
  {\bibfnamefont {R.~D.}\ \bibnamefont {{Hoffman}}}, \bibinfo {author}
  {\bibfnamefont {T.}~\bibnamefont {{Rauscher}}}, \bibinfo {author}
  {\bibfnamefont {A.}~\bibnamefont {{Sakharuk}}}, \bibinfo {author}
  {\bibfnamefont {H.}~\bibnamefont {{Schatz}}}, \bibinfo {author}
  {\bibfnamefont {F.~K.}\ \bibnamefont {{Thielemann}}},\ and\ \bibinfo {author}
  {\bibfnamefont {M.}~\bibnamefont {{Wiescher}}},\ }\href
  {https://doi.org/10.1088/0067-0049/189/1/240} {\bibfield  {journal} {\bibinfo
   {journal} {Astrophys. J. Suppl. Ser.}\ }\textbf {\bibinfo {volume} {189}},\
  \bibinfo {pages} {240} (\bibinfo {year} {2010})}\BibitemShut {NoStop}%
\bibitem [{Note1()}]{Note1}%
  \BibitemOpen
  \bibinfo {note} {As Ref.~\cite {bliss2020} shows, the impact of an individual
  reaction rate depends on the rates adopted for many nuclear reactions. For
  our estimated impact, we are concentrating on the average linear trend of
  their silver abundance versus $^{96}{\protect \rm Zr}(\alpha ,n)$ rate
  variation scatter plot.}\BibitemShut {Stop}%
\bibitem [{\citenamefont {{Roederer}}\ \emph {et~al.}(2012)\citenamefont
  {{Roederer}}, \citenamefont {{Lawler}}, \citenamefont {{Sobeck}},
  \citenamefont {{Beers}}, \citenamefont {{Cowan}}, \citenamefont {{Frebel}},
  \citenamefont {{Ivans}}, \citenamefont {{Schatz}}, \citenamefont {{Sneden}},\
  and\ \citenamefont {{Thompson}}}]{Roed12}%
  \BibitemOpen
  \bibfield  {author} {\bibinfo {author} {\bibfnamefont {I.~U.}\ \bibnamefont
  {{Roederer}}}, \bibinfo {author} {\bibfnamefont {J.~E.}\ \bibnamefont
  {{Lawler}}}, \bibinfo {author} {\bibfnamefont {J.~S.}\ \bibnamefont
  {{Sobeck}}}, \bibinfo {author} {\bibfnamefont {T.~C.}\ \bibnamefont
  {{Beers}}}, \bibinfo {author} {\bibfnamefont {J.~J.}\ \bibnamefont
  {{Cowan}}}, \bibinfo {author} {\bibfnamefont {A.}~\bibnamefont {{Frebel}}},
  \bibinfo {author} {\bibfnamefont {I.~I.}\ \bibnamefont {{Ivans}}}, \bibinfo
  {author} {\bibfnamefont {H.}~\bibnamefont {{Schatz}}}, \bibinfo {author}
  {\bibfnamefont {C.}~\bibnamefont {{Sneden}}},\ and\ \bibinfo {author}
  {\bibfnamefont {I.~B.}\ \bibnamefont {{Thompson}}},\ }\href
  {https://doi.org/10.1088/0067-0049/203/2/27} {\bibfield  {journal} {\bibinfo
  {journal} {Astrophys. J. Suppl. Ser.}\ }\textbf {\bibinfo {volume} {203}},\
  \bibinfo {eid} {27} (\bibinfo {year} {2012})}\BibitemShut {NoStop}%
\bibitem [{\citenamefont {{Hansen}}\ \emph {et~al.}(2012)\citenamefont
  {{Hansen}}, \citenamefont {{Primas}}, \citenamefont {{Hartman}},
  \citenamefont {{Kratz}}, \citenamefont {{Wanajo}}, \citenamefont
  {{Leibundgut}}, \citenamefont {{Farouqi}}, \citenamefont {{Hallmann}},
  \citenamefont {{Christlieb}},\ and\ \citenamefont {{Nilsson}}}]{Hans12}%
  \BibitemOpen
  \bibfield  {author} {\bibinfo {author} {\bibfnamefont {C.~J.}\ \bibnamefont
  {{Hansen}}}, \bibinfo {author} {\bibfnamefont {F.}~\bibnamefont {{Primas}}},
  \bibinfo {author} {\bibfnamefont {H.}~\bibnamefont {{Hartman}}}, \bibinfo
  {author} {\bibfnamefont {K.~L.}\ \bibnamefont {{Kratz}}}, \bibinfo {author}
  {\bibfnamefont {S.}~\bibnamefont {{Wanajo}}}, \bibinfo {author}
  {\bibfnamefont {B.}~\bibnamefont {{Leibundgut}}}, \bibinfo {author}
  {\bibfnamefont {K.}~\bibnamefont {{Farouqi}}}, \bibinfo {author}
  {\bibfnamefont {O.}~\bibnamefont {{Hallmann}}}, \bibinfo {author}
  {\bibfnamefont {N.}~\bibnamefont {{Christlieb}}},\ and\ \bibinfo {author}
  {\bibfnamefont {H.}~\bibnamefont {{Nilsson}}},\ }\href
  {https://doi.org/10.1051/0004-6361/201118643} {\bibfield  {journal} {\bibinfo
   {journal} {Astron. \& Astrophys.}\ }\textbf {\bibinfo {volume} {545}},\
  \bibinfo {eid} {A31} (\bibinfo {year} {2012})}\BibitemShut {NoStop}%
\bibitem [{\citenamefont {{Lamere}}\ \emph {et~al.}(2019)\citenamefont
  {{Lamere}}, \citenamefont {{Couder}}, \citenamefont {{Beard}}, \citenamefont
  {{Simon}}, \citenamefont {{Simonetti}}, \citenamefont {{Skulski}},
  \citenamefont {{Seymour}}, \citenamefont {{Huestis}}, \citenamefont
  {{Manukyan}}, \citenamefont {{Meisel}}, \citenamefont {{Morales}},
  \citenamefont {{Moran}}, \citenamefont {{Moylan}}, \citenamefont
  {{Seymour}},\ and\ \citenamefont {{Stech}}}]{Lame19}%
  \BibitemOpen
  \bibfield  {author} {\bibinfo {author} {\bibfnamefont {E.}~\bibnamefont
  {{Lamere}}}, \bibinfo {author} {\bibfnamefont {M.}~\bibnamefont {{Couder}}},
  \bibinfo {author} {\bibfnamefont {M.}~\bibnamefont {{Beard}}}, \bibinfo
  {author} {\bibfnamefont {A.}~\bibnamefont {{Simon}}}, \bibinfo {author}
  {\bibfnamefont {A.}~\bibnamefont {{Simonetti}}}, \bibinfo {author}
  {\bibfnamefont {M.}~\bibnamefont {{Skulski}}}, \bibinfo {author}
  {\bibfnamefont {G.}~\bibnamefont {{Seymour}}}, \bibinfo {author}
  {\bibfnamefont {P.}~\bibnamefont {{Huestis}}}, \bibinfo {author}
  {\bibfnamefont {K.}~\bibnamefont {{Manukyan}}}, \bibinfo {author}
  {\bibfnamefont {Z.}~\bibnamefont {{Meisel}}}, \bibinfo {author}
  {\bibfnamefont {L.}~\bibnamefont {{Morales}}}, \bibinfo {author}
  {\bibfnamefont {M.}~\bibnamefont {{Moran}}}, \bibinfo {author} {\bibfnamefont
  {S.}~\bibnamefont {{Moylan}}}, \bibinfo {author} {\bibfnamefont
  {C.}~\bibnamefont {{Seymour}}},\ and\ \bibinfo {author} {\bibfnamefont
  {E.}~\bibnamefont {{Stech}}},\ }\href
  {https://doi.org/10.1103/PhysRevC.100.034614} {\bibfield  {journal} {\bibinfo
   {journal} {\prc}\ }\textbf {\bibinfo {volume} {100}},\ \bibinfo {eid}
  {034614} (\bibinfo {year} {2019})}\BibitemShut {NoStop}%
\bibitem [{\citenamefont {{Montanari}}\ and\ \citenamefont
  {{Dimitriou}}(2017)}]{Mont17}%
  \BibitemOpen
  \bibfield  {author} {\bibinfo {author} {\bibfnamefont {C.~C.}\ \bibnamefont
  {{Montanari}}}\ and\ \bibinfo {author} {\bibfnamefont {P.}~\bibnamefont
  {{Dimitriou}}},\ }\href {https://doi.org/10.1016/j.nimb.2017.03.138}
  {\bibfield  {journal} {\bibinfo  {journal} {Nucl. Instrum. Meth. Phys. Res.
  Sec. B}\ }\textbf {\bibinfo {volume} {408}},\ \bibinfo {pages} {50} (\bibinfo
  {year} {2017})}\BibitemShut {NoStop}%
\bibitem [{\citenamefont {{Baglin}}(2011)}]{Bagl11}%
  \BibitemOpen
  \bibfield  {author} {\bibinfo {author} {\bibfnamefont {C.~M.}\ \bibnamefont
  {{Baglin}}},\ }\href {https://doi.org/10.1016/j.nds.2011.04.001} {\bibfield
  {journal} {\bibinfo  {journal} {Nucl. Data Sheets}\ }\textbf {\bibinfo
  {volume} {112}},\ \bibinfo {pages} {1163} (\bibinfo {year}
  {2011})}\BibitemShut {NoStop}%
\bibitem [{\citenamefont {{Abriola}}\ and\ \citenamefont
  {{Sonzogni}}(2006)}]{Abri06}%
  \BibitemOpen
  \bibfield  {author} {\bibinfo {author} {\bibfnamefont {D.}~\bibnamefont
  {{Abriola}}}\ and\ \bibinfo {author} {\bibfnamefont {A.~A.}\ \bibnamefont
  {{Sonzogni}}},\ }\href {https://doi.org/10.1016/j.nds.2006.08.001} {\bibfield
   {journal} {\bibinfo  {journal} {Nucl. Data Sheets}\ }\textbf {\bibinfo
  {volume} {107}},\ \bibinfo {pages} {2423} (\bibinfo {year}
  {2006})}\BibitemShut {NoStop}%
\bibitem [{\citenamefont {{Nica}}(2010)}]{Nica10}%
  \BibitemOpen
  \bibfield  {author} {\bibinfo {author} {\bibfnamefont {N.}~\bibnamefont
  {{Nica}}},\ }\href {https://doi.org/10.1016/j.nds.2010.03.001} {\bibfield
  {journal} {\bibinfo  {journal} {Nucl. Data Sheets}\ }\textbf {\bibinfo
  {volume} {111}},\ \bibinfo {pages} {525} (\bibinfo {year}
  {2010})}\BibitemShut {NoStop}%
\end{thebibliography}%
\end{document}